%% file: main.tex
\documentclass[lettersize,journal]{IEEEtran}

\usepackage{amsmath,amsfonts,amssymb}
\usepackage{bbm}
\usepackage{dsfont}
\usepackage{nicefrac}

\usepackage{multirow}
\usepackage{multicol}
\usepackage{booktabs}
\usepackage{makecell}
\usepackage{tabularx}
\usepackage{adjustbox}

\usepackage{graphicx}
\usepackage{subfig}   
\usepackage{float}

\usepackage{algpseudocode}
\usepackage[ruled,vlined,linesnumbered,lined,boxed,commentsnumbered,ruled,longend,noend]{algorithm2e}

\usepackage{xcolor}

\usepackage{cite}

\usepackage[nolist]{acronym}
\usepackage{comment}
\usepackage{fancyhdr,lipsum}

\fancypagestyle{mahmood}{%
   \fancyhf{} 
   
   \fancyhead[C]{\tiny This manuscript has been accepted for publication in \textbf{IEEE Transactions on Network and Service Management}. 
   
   You can use this material personally. Reprinting or republishing this material for the purpose of advertising or promotion, creating new collective works, 
   
   reselling or redistributing to servers or lists, or using any copyrighted component in other works \textbf{must adhere to IEEE policy}. DOI: TBA.}
}%

\newcommand\myeq{\mathrel{\stackrel{\makebox[0pt]{\mbox{\normalfont\tiny (1), (2)}}}{=}}}

\begin{document}
\title{A Semantic-Aware Multiple Access Scheme Leveraging Spatial Redundancy for Uplink-Dominant Network Services}

\author{
    \IEEEauthorblockN{
        Hamidreza Mazandarani\textsuperscript{1}, Masoud Shokrnezhad\textsuperscript{2}, and Tarik Taleb\textsuperscript{1} \\
    }
    \IEEEauthorblockA{
       \textsuperscript{1} \textit{Ruhr University Bochum, Bochum, Germany; \{hamidreza.mazandarani, tarik.taleb\}@rub.de} \\
        \textsuperscript{2} \textit{ICTFicial Oy, Espoo, Finland; masoud.shokrnezhad@ictficial.com}
    }
}

\maketitle
\thispagestyle{mahmood}

\begin{abstract}
The transition toward semantic-aware communication offers a paradigm shift for \textcolor{black}{next-generation mobile} networks, promising to decouple information significance from raw data transmission. Despite advances in semantic extraction, the integration of semantic intelligence into the Medium Access Control (MAC) layer remains underexplored, particularly in exploiting spatial correlations among users. To address this, we introduce a novel multiple access scheme designed for \textcolor{black}{uplink-dominant network services}. This framework optimizes the trade-off between spectrum utilization and sustainability by formulating variable-packet-length access as distinct $\alpha$-fairness and energy efficiency problems. A key innovation of our approach is the quantification of spatial redundancies through novel metrics of self-throughput and assisted-throughput, which account for the semantic correlation of data across user equipment. We analyze these formulations to identify optimal bounds before proposing PRISM (Protocol for Redundancy Identification in Semantic Multiple-access). Grounded in Model-free Multi-Agent Deep Reinforcement Learning (MADRL), PRISM enables devices to autonomously govern spectrum access using only local observations. Extensive evaluations demonstrate that PRISM successfully leverages redundancies to outperform semantic-oblivious schemes, achieving \textcolor{black}{up to \({90\%}\) of the centralized optimal benchmark and improving both objectives by up to \({2\times}\)} across diverse user-semantic association matrices. These results validate PRISM as a viable candidate for future distributed \textcolor{black}{mobile network} applications, complemented by orthogonal Multiple Access Schemes where signals are multiplexed in the semantic domain.

\end{abstract}


\input{sections/1_introduction}

\input{sections/2_background}
\input{sections/3_formulation}
\input{sections/4_PRISM}
\input{sections/5_evaluation}
\input{sections/6_conclusion}

\section*{Acknowledgment}
The research work is supported in part by the Federal Ministry of Research, Technology, and Space (BMFTR), Germany, through the Project 6GEM+ under Grant 16KIS2411; and in part by the European Union's Horizon Europe research and innovation programme under the 6G-Path project (Grant No. 101139172).

\bibliographystyle{IEEEtran}
\bibliography{IEEEabrv,main}

\end{document}

%% file: sections/1_introduction.tex
\section{Introduction}\label{S_INT}

\begin{figure*}[!t]
\centerline{\includegraphics[width=7in]{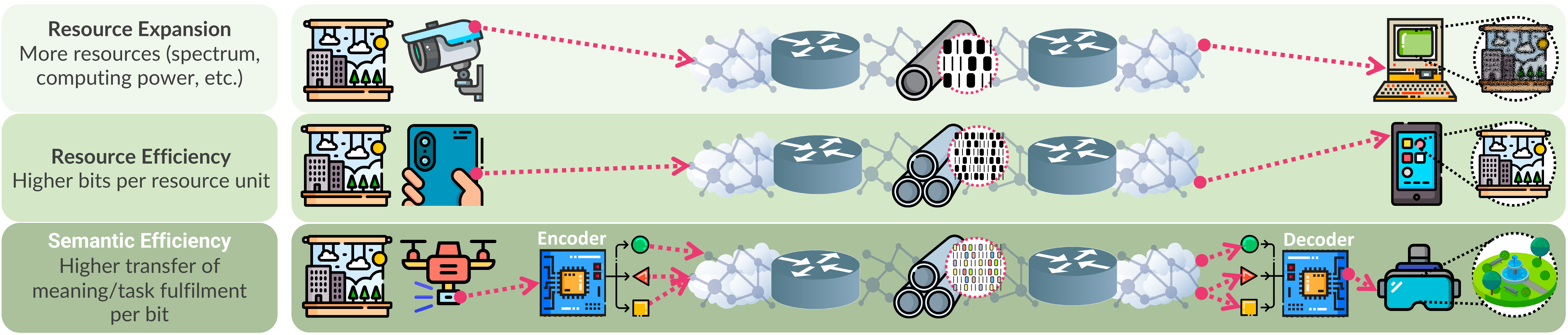}}
\caption{The emergence of the beyond-Shannon paradigm evolved from unsustainable resource-expansion mindsets (e.g., increasing spectrum bands) and theoretically limited resource-efficiency improvement methods (e.g., novel multiplexing techniques) to enhancing semantic efficiency.}
\label{vision}
\end{figure*}

\begin{figure}[t!]\centering
\centerline{\includegraphics[width=3in]{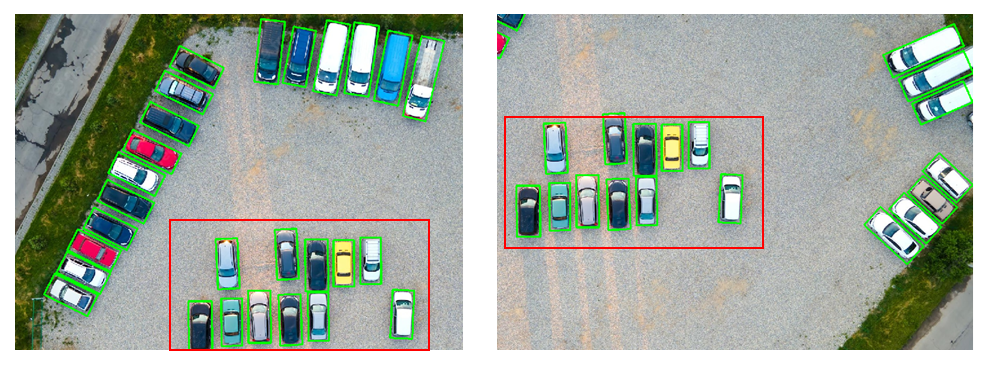}}
\caption{Two UAVs with overlapping observations of a parking lot, sharing common semantic segments (i.e., vehicles within the red rectangular area), creating an opportunity to avoid excessive resource usage while preserving the system’s cognitive performance. Image taken from the DOTA dataset \cite{xia2018dota}.}
\label{illustrative_example}
\end{figure}

Communication systems are transitioning from a traditional \textit{bit-oriented} mindset, which focuses solely on transmitting the maximum number of data bits by either expanding resources or enhancing resource efficiency, to a \textit{semantic-aware} paradigm. In this novel approach, the semantic value of various system components (including messages, users, and resources) is quantified and subsequently integrated into the service provisioning process to enhance semantic efficiency (Fig. \ref{vision}) \cite{our_mag_paper}. This paradigm shift opens up significant opportunities for innovative services, particularly within the context of the 6G-based Metaverse, where the seamless integration of physical and virtual environments is a primary objective \cite{haoyu2023xr}. Furthermore, such a minimalist strategy, based on the design philosophy of \textit{Less data, more knowledge} \cite{chaccour2022less}, promotes the development of sustainable networks \cite{agheli2024semantic} by prioritizing the efficient and equitable use of resources. As an illustrative example, consider Unmanned Aerial Vehicles (UAVs) transmitting partially overlapping observations to a central server. By identifying semantic segments such as vehicles and avoiding repetitive transmissions, the system’s cognitive performance is preserved with reduced bandwidth and energy usage (Fig. \ref{illustrative_example}).


While semantic awareness has been explored across various wireless domains, \textcolor{black}{including adaptive video streaming \cite{10973307} and channel identification \cite{10820115}}, its application at the data link layer, particularly within the Medium Access Control (\acs{MAC}) protocol, still requires further investigation. This direction is promising, as it offers significant potential to improve network efficiency through more intelligent management of data packet transmissions and resource allocation \cite{our_mag_paper, lin2023meta, chaccour2022less, strinati2024goal}. Early examples, such as the Model Division Multiple Access (\acs{MDMA}), demonstrate the promise of this approach by allocating resources for shared information transmission among users \cite{zhang2023model}. Another research direction involves joint semantic encoding for multiple users, known as multi-user semantic communications \cite{xie2022task,zhang2023deepma}, although these methods often lack adaptability and scalability for 6G systems. Conversely, Multi-Agent Deep Reinforcement Learning (\acs{MADRL}) is effective in environments with multiple evolving actors \cite{gronauer2022multi}. MADRL adapts to non-stationarity by making transmission decisions based on historical data and coordinating through policy sharing and centralized training. Despite its potential \cite{naparstek2018deep, Sohaib2021dynamic, guo2022multi}, most current MADRL-based multiple access controls do not incorporate semantic context understanding, which is increasingly necessary for the evolving communication landscape.

In response to the existing gap in the current literature, we have recently introduced a novel approach to address the semantic-aware, distributed multiple-access challenge \cite{wcnc_2024}. Referred to as the Semantic-Aware Multi-Agent Double and Dueling Deep Q-Learning (SAMA-D3QL) method and expanded upon the established principles outlined \textcolor{black}{in our prior investigations \cite{TMLCN_2024,mazandarani2023self, mazandarani2025semantic}}, our method aims to optimize the trade-off between resource utilization and fairness, taking into account the inherent correlation present in users' data. It operates on the assumption that users engage in the transmission of diverse semantic segments, with certain segments being potentially shared among them. Consequently, when a user initiates transmission, it confers benefits upon all users with shared relevant segments. Drawing upon the insights garnered from SAMA-D3QL, this study offers noteworthy contributions to the field, as outlined below:

\begin{itemize}
    \item We introduce a comprehensive framework in which users' utilities are not as independent as is commonly assumed in traditional multiple access networks. This interdependence arises from the existence of shared segments among certain subsets of users, where the transmissions of one user can impact the task fulfillment of others. Examples of such interdependence are observed in scenarios involving holographic presence, distributed Generative Artificial Intelligence (GenAI), and multi-UAV area coverage \cite{mazandarani2025semantic}.
    \item In pursuit of environmental and societal sustainability within the distributed multiple access capability of 6G networks, we leverage energy efficiency as an objective function, in addition to the fairness-utilization trade-off. Our energy efficiency assessment includes a parameterized consideration of the energy consumption associated with semantic encoding, thereby offering a more precise and realistic understanding of the system and aiming to reduce the environmental footprint of the network.
    \item We formulate the challenge as a novel optimization problem for each objective function, incorporating the variable packet length of User Equipment (\acs{UE}) and capturing semantic correlations within the resulting asynchronous setup.
    \item We perform a comparative analysis of optimal solutions for each problem concerning its objective function: energy efficiency, quantified as the difference between users' throughput and coefficient-weighted consumed energy, and the fairness-utilization trade-off, evaluated using alpha-fairness metrics with varying values of $\alpha$. While achieving optimal solutions in real-world scenarios is computationally infeasible due to the complexity of the problems, these solutions serve as target bounds for environmental and societal sustainability in the design of real-time approaches.
    \item We develop a MADRL-based approach, named PRISM, to solve these problems in a practical manner by incorporating a novel observation space and reward mechanism tailored to semantics and asynchrony. Specifically, we employ only fully-connected layers as the approximation function in the MADRL algorithm, processing short- and long-term histories with an averaged context vector, thereby enhancing the system's sustainability. Moreover, we design objective-specific reward shaping to achieve the right trade-offs for both the $\alpha$-fairness and energy efficiency objectives.
\end{itemize}

The subsequent sections of this paper are structured as follows: Section {\ref{S_B&RW}} provides the necessary background and a comprehensive literature review. Section {\ref{S_PA}} explains the motivation behind this work, introduces the system model, and presents a thorough problem formulation. Additionally, this section proposes an interpretation of the optimal solutions. Section {\ref{S_SES}} elaborates on PRISM, detailing its learning mechanism and training algorithm. Following this, Section {\ref{S_EVA}} presents an exposition of the numerical results obtained. Finally, Section {\ref{S_CON}} concludes our work with closing remarks on our findings.


\input{acro_list.tex}

%% file: acro_list.tex
\begin{acronym}[MPC] 
\acro{MADRL}{Multi-Agent Deep Reinforcement Learning}
\acro{MAC}{Medium Access Control}
\acro{MDMA}{Model Division Multiple Access}
\acro{SemCom}{Semantic Communications}
\acro{UE}{User Equipment}
\acro{UAV}{Unmanned Aerial Vehicles}
\end{acronym}

%% file: sections/2_background.tex
\section{Background and Related Work}\label{S_B&RW}

\subsection{From Bits to Semantics}

Historically, the response to emerging applications in wireless communications has predominantly focused on expanding resources (e.g., exploring new spectrum bands) or enhancing resource efficiency (e.g., advancing MIMO technology). Nonetheless, these efforts are insufficient to support innovative communication services, particularly the 6G-based Metaverse, where seamless integration of physical and virtual environments is a primary goal \cite{haoyu2023xr}. Fortunately, a semantic revolution is underway, shifting from bit transmission to the transfer of semantics among transceivers, along with semantic-aware resource allocation and orchestration. Semantic Communications (SemCom) aims for the receiver to accurately interpret the transmitted message as intended by the sender \cite{nguyen2025contemporary}. This objective can be realized differently based on system goals and data modalities. For example, the transmitter may identify and prioritize transmission of key segments in images via \textcolor{black}{segmentation models like the Segment Anything Model (SAM) versions one and two \cite{ravi2024sam}}. Moreover, joint source and channel coding can be applied in digital communications to transfer semantics despite channel impairments \cite{park2024joint}. Graph-semantic compression is also another option to realize SemCom \cite{sangaiah2024r}. Nonetheless, SemCom necessitates mutual understanding, or even a theory of mind \cite{nguyen2025contemporary}, among transceivers, which involves an awareness of the entire communicative system regarding semantics, to be discussed in the next subsection.

\subsection{Semantic-Awareness Beyond the Physical Layer}

The integration of semantics into decision-making processes at various network layers remains in its early stages of development. Our knowledge management framework for semantic-aware systems, termed \textit{Knowledge Base Management and Orchestration} (KB MANO) \cite{our_mag_paper}, alongside the concept of the knowledge space within the Meta-Networking framework \cite{lin2023meta}, the reasoning plane proposed by Lin \textit{et al.}~\cite{chaccour2022less}, and the 6G-GOALS architecture \cite{strinati2024goal}, represent some of the initial efforts to conceptualize network-wide semantic awareness. Additionally, there is an increasing focus on specific semantic-aware functionalities, such as routing, radio resource allocation, and service provisioning, as surveyed by Getu \textit{et al.}~\cite{getu2024survey} and Trevlakis \textit{et al.}~\cite{trevlakis2024towards}. Among these, the enhancement of multiple access to the frequency spectrum through semantic awareness holds significant potential. As the initial point where user data enters the external environment, the MAC layer can significantly benefit from semantic-aware protocols by eliminating redundant transmissions and optimizing resource use.

MDMA is an early example of exploiting this potential, dedicating a portion of resources to transmit shared information among users \cite{zhang2023model}. Orthogonal MDMA (O-MDMA) further advances this scheme by introducing the concept of semantic orthogonal signals, i.e., the capability to treat signals generated by different semantic encoders as interference while meeting user performance requirements \cite{liang2024orthogonal}. O-MDMA also accommodates multiple modalities and variable semantic signal lengths. Similarly, Kassab \textit{et al.}~\cite{kassab2020multi} explored scenarios involving a collective of Internet of Things (IoT) devices monitoring correlated on-off events, such as temperature variations. The authors suggest that devices observing the same event could send redundant information, so they propose maximizing the average total event rate of the system instead of the traditional frame throughput. More recently, Qiao \textit{et al.}~\cite{qiao2025token} propose token-domain multiple access (ToDMA), which leverages the semantic orthogonality of tokens. These tokens result from mapping input segments into discrete information units using techniques like vector quantization. 

Another research direction, exemplified by DeepSC-IR/MT \cite{xie2022task} and DeepMA \cite{zhang2023deepma}, investigates the joint semantic encoding of data from multiple users, also known as multi-user semantic communications. Although these methods are effective in certain contexts, they do not adequately address the distributed nature of future applications with shared user information, limiting their scalability and suitability for 6G systems. In this regard, advanced decision-making algorithms, such as RL, could elevate semantic awareness to a higher level.


\subsection{MADRL: An enabler for semantic-aware Multiple Access}

In environments characterized by ever-changing, multiple distributed actors, such as unsourced multiple access in 6G \cite{agostini2024evolution} or UAV-assisted wireless sensor networks \cite{10664472}, MADRL proves to be an effective approach. Its adaptability to handle non-stationarity, resulting from the co-evolution of agents’ policies, underscores its suitability for such decentralized contexts \cite{gronauer2022multi}. In MADRL frameworks, agents make transmission decisions based on historical observations, including the success or failure of task completions (e.g., successful transmissions in the multiple access problem), while coordination among agents is facilitated through mechanisms such as policy sharing, message passing, or centralized training.

MADRL-based multiple access control has been discussed in various papers. For instance, Naparstek \textit{et al.}~\cite{naparstek2018deep} proposed an efficient Nash equilibrium for multi-agent MAC where all agents adopt the same policy. Sohaib \textit{et al.}~\cite{Sohaib2021dynamic} extended MADRL to MAC by using transmission policies over consecutive time slots as decision variables (as opposed to per-time-slot decision-making), ensuring short-term fairness under varying user conditions. Guo \textit{et al.}~\cite{guo2022multi} introduced the QMIX-advanced LBT (QLBT) MAC protocol, building upon the QMIX algorithm for MADRL problems \cite{rashid2020monotonic}. QLBT incorporates two reward signals, one for maximizing network utilization and another for maintaining fairness. Miuccio \textit{et al.}~\cite{miuccio2024learning} designed a MAC algorithm considering communication overhead issues (e.g., scheduling requests and grants) with generalization capability using auto-encoder-based state abstraction. However, despite these advancements, none of the examined MADRL-based MAC protocols explicitly address the growing need for semantic awareness, a critical aspect of evolving communication systems.

\subsection{Sustainable Semantic-Awareness: the final puzzle piece}

Alongside being distributed, any proposed approach for future systems should consider energy efficiency as a critical objective for optimization. Energy efficiency is regarded as a key factor in environmental sustainability, also impacting business and economic sustainability \cite{ahmadi2025towards}. The minimalist design approach of semantic communications paves the way to address this challenge towards sustainable networks \cite{agheli2024semantic}. The promises of semantic-aware network orchestration could be even more transformative, given its influence on the entire communication system. This approach allows the network to prioritize essential and high-value information, thereby reducing the need to transmit redundant or low-priority data. By minimizing unnecessary transmissions, the overall energy usage of the network is significantly decreased. Additionally, semantic-aware systems can facilitate more efficient resource allocation and management by sharing or deactivating a substantial portion of resources, thereby reducing energy waste in network operations and maintenance. Conversely, the energy consumption associated with semantic learning and extraction could negatively impact sustainability.

The energy-defined networking concept, which integrates energy measurement components across all system layers \cite{galis2024future}, offers a viable solution by allowing the consideration of energy consumption in evaluating system efficiency and sustainability. Despite this, only a limited number of studies address the sustainability aspects of semantic communications. Some research targets the computational disparities between the capabilities of different users \cite{nguyen2024swin, albaseer2024tailoring}, while others focus on the efficiency of semantic communications training \cite{nguyen2024efficient}. However, it is necessary to consider more precise energy consumption models and investigate the impact of practical metrics for real-world usage, such as semantic encoding, on system efficiency and effectiveness, drawing inspiration from the work on incorporating decision-making costs into energy efficiency in multiple access \cite{mazandarani2021energy}.

\subsection{Our Approach}

Complementary to some existing work that concentrates on the semantic \textit{orthogonality} of users \cite{zhang2023model, liang2024orthogonal, qiao2025token}, we target \textit{similarities} among users' information. To this end, we propose a comprehensive framework that challenges the conventional assumption of independent user utilities, considering the interdependence arising from shared semantic segments among user subsets. Such interdependence is relevant in scenarios like holographic presence, distributed Generative AI, and multi-UAV area coverage. \textcolor{black}{Exploiting redundancy among users is not new in itself: correlation-based MAC allows a representative node to report on behalf of its correlation neighbors \cite{vuran2006spatial}, redundant coverage can be provisioned at deployment time in sensor networks \cite{11278102}, and event-driven access rewards the events reported rather than the frames delivered \cite{kassab2020multi}. In these schemes, the redundancy structure is given externally, through deployment geometry, a commonly observed event, or a model shared across encoders, and is therefore known before access decisions are made. Here, in contrast, redundancy is content-defined: which semantic segments are shared, and by which subset of UEs, changes over time and is not known at the UEs. Each UE must instead infer its semantic contribution from access feedback and act on that information in a distributed, asynchronous manner.}

Our work incorporates energy efficiency, factoring in the energy costs of semantic encoding as a key objective alongside the fairness-utilization trade-off. The problem is formulated as a novel optimization task that accounts for variable packet lengths and semantic correlations in an asynchronous setting. Here, variable packet length refers to transmitting packets that span more than one frame, highlighting the system’s greater freedom in decision-making \cite{9079169, TMLCN_2024}. In Table \ref{tab_paper_comparison}, a comparison of the studied papers about MAC design, including their main ideas and characteristics, is presented.

\begin{table*}[!hbt]
\caption{Selected Multiple Access Schemes Comparison.}
\centering
\begin{adjustbox}{width=1\textwidth}
\begin{tabular}{|l|l|l|c|c|}
\hline
\multicolumn{1}{|c|}{\textbf{Reference}} &
  \multicolumn{1}{c|}{\textbf{Main Idea}} &
  \multicolumn{1}{c|}{\textbf{Objective Function}} &
  \textbf{\begin{tabular}[c]{@{}c@{}}Semantic\\ Awareness\end{tabular}} &
  \textbf{\begin{tabular}[c]{@{}c@{}}Time\\ Awareness\end{tabular}} \\ \hline
MDMA \cite{zhang2023model} &
  \begin{tabular}[c]{@{}l@{}}Dedicating a portion of resources\\ to transmit shared information among users\end{tabular} &
  Peak signal-to-noise ratio &
  \checkmark &
  --- \\ \hline
O-MDMA \cite{liang2024orthogonal} &
  \begin{tabular}[c]{@{}l@{}}Semantic orthogonal signals, \\ i.e. the capability to treat signals\\ generated by different semantic encoders as \\  interference while meeting user performance \\ requirements \end{tabular} &
  Multi-scale structural similarity &
  \checkmark &
  --- \\ \hline
Kassab \textit{et al.}~\cite{kassab2020multi} &
  \begin{tabular}[c]{@{}l@{}}Avoid transmitting redundant information\\ by IoT devices observing the same event\end{tabular} &
  Average sum event rate &
  \checkmark &
  \checkmark \\ \hline
DeepSC-IR/MT \cite{xie2022task} &
  \begin{tabular}[c]{@{}l@{}}A Transformer-based transmitter structure\\ to extract semantic information for different tasks,\\ while eliminating interference from other users\end{tabular} &
  \begin{tabular}[c]{@{}l@{}}- Image: Recall@1\\ - Text: Bilingual evaluation understudy\\ - Video: Answer accuracy\end{tabular} &
  \checkmark &
  --- \\ \hline
DeepMA \cite{zhang2023deepma} &
  \begin{tabular}[c]{@{}l@{}}Transmitting multiple data instances simultaneously\\ over a shared communication channel\\ via training the network to recover the target data\\ from a mixed semantic symbol vector\end{tabular} &
  Peak signal-to-noise ratio &
  \checkmark &
  \checkmark \\ \hline
Naparstek \textit{et al.}~\cite{naparstek2018deep} &
  \begin{tabular}[c]{@{}l@{}}Adopt MADRL for distributed dynamic \\ spectrum access + game theory analysis\end{tabular} &
  Average rate &
  --- &
  \checkmark \\ \hline
Sohaib \textit{et al.}~\cite{Sohaib2021dynamic} &
  \begin{tabular}[c]{@{}l@{}}Selecting transmission policies over consecutive\\ time slots ensuring short-term fairness \\under varying user conditions\end{tabular} &
  \begin{tabular}[c]{@{}l@{}}- Throughput\\ - Fairness\end{tabular} &
  --- &
  \checkmark \\ \hline
QLBT \cite{guo2022multi} &
  \begin{tabular}[c]{@{}l@{}}Leveraging QMIX MADRL algorithm, in addition \\ to novel observations and reward mechanism\end{tabular} &
  \begin{tabular}[c]{@{}l@{}}- Throughput\\ - Fairness\end{tabular} &
  --- &
  \checkmark \\ \hline
Miuccio \textit{et al.}~\cite{miuccio2024learning} &
  \begin{tabular}[c]{@{}l@{}}Considering communication overhead issues\\ with generalization capability\\ using auto-encoder-based state abstraction\end{tabular} &
  \begin{tabular}[c]{@{}l@{}}- Throughput\\ - Generalizability\end{tabular} &
  --- &
  \checkmark \\ \hline
\begin{tabular}[c]{@{}l@{}}SAMA-D3QL \cite{wcnc_2024}\\ (Our previous work)\end{tabular} &
  \begin{tabular}[c]{@{}l@{}}Accounting for user data correlation\\ by introducing self- and assisted throughputs\\ in MADRL-based MAC scheme design\end{tabular} &
  ${\alpha}$-fairness &
  \checkmark &
  \checkmark \\ \hline
PRISM (This work) &
  \begin{tabular}[c]{@{}l@{}}Expanding SAMA-D3QL to variable packet \\ length setting, and sustainability awareness\end{tabular} &
  \begin{tabular}[c]{@{}l@{}}- ${\alpha}$-fairness\\ - Energy efficiency\end{tabular} &
  \checkmark &
  \checkmark \\ \hline
\end{tabular}
\end{adjustbox}
\label{tab_paper_comparison}
\end{table*}

%% file: sections/3_formulation.tex
\section{System model and Problem Formulation}\label{S_PA}

\subsection{System Model}

An indoor area comprising a Small Base Station (SBS) and a collection of $\mathcal{N}$ intelligent User Equipment (UE) is considered, each uniquely labeled as $u_{i}$, where $i \in \mathbb{N} = \{1, \ldots, \mathcal{N}\}$. These UEs are engaged in contention for access to $\mathcal{C}$ perfectly time-slotted communication channels designated for uploading their data to the SBS. It is essential to note that simultaneous transmissions over a single channel lead to a collision. Consequently, in the conventional bit-oriented framework, the highest achievable network utilization is capped at $\mathcal{C}$ packets per time slot.

\begin{figure}[t!]\centering
\includegraphics[width=2.2in]{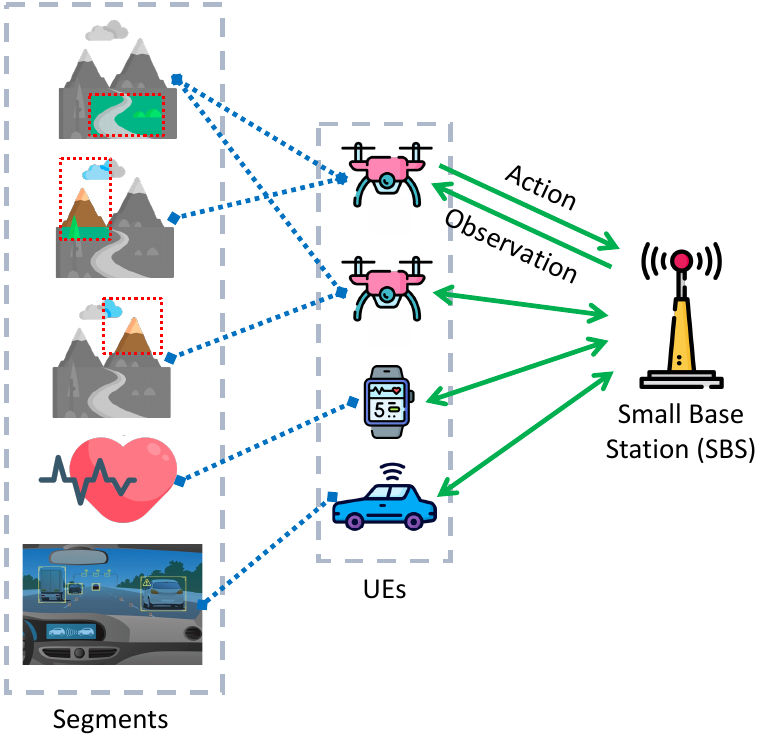}
    \vspace{-5pt}
    \caption{A sample network consisting of four UEs ($\mathcal{N} = 4$) and five segments ($\mathcal{K} = 5$). As illustrated, one separate segment is associated with each user, in addition to one shared segment among the first and second UEs.}
    \vspace{-15pt}
    \label{fig_sample_net}
\end{figure}

The semantic space is represented by $\mathcal{K}$ distinct entities, termed \textit{segments}, denoted as $k \in \mathbb{K} = \{1, \ldots, \mathcal{K} \}$. At each time slot $t \in \{0, \ldots, \mathcal{T} \}$, each user $u_{i}$ is associated/disassociated with segment $k$, signified by a binary variable $a_{i,k}^{t} \in \{0, 1\}$, forming a predefined binary association matrix $\mathbb{A} = [a_{i,k}^{t}]_{\mathcal{N} \times \mathcal{K} \times \mathcal{T}}$. \textcolor{black}{Operationally, $a_{i,k}^{t}=1$ means that the semantic descriptor extracted by UE $i$ at slot $t$ is assigned to the global segment $k$ according to a service-specific matching rule, such as label agreement, spatial registration, hash matching, or embedding similarity.} The matrix serves as the input to the system, and an orchestration entity, such as KB MANO as discussed by Shokrnezhad \textit{et al.} \cite{our_mag_paper}, is responsible for generating it. In an illustrative example, shown in Fig. \ref{fig_sample_net}, the association matrix at time slot $t$ is given as $\mathbb{A}^{t} = \big[[1, 1, 0, 0, 0],[1, 0, 1, 0, 0],[0, 0, 0, 1, 0],[0, 0, 0, 0, 1]\big]$. The following subsection elaborates further on the details of these semantic correlations and their impact on the system performance.

\vspace{3 pt}

\subsubsection{Semantic Spatial Correlations}

To evaluate the semantic efficiency of the system, we introduce a \textit{segment transmission} indicator, denoted as $y_{k}^{t} \in \{0, 1\}$. This indicator operates under the assumption that, at any given time slot, all UE-associated segments are combined into a single packet. Furthermore, receiving a specific segment by the SBS multiple times within a time slot does not enhance throughput. $y_{k}^{t}$ is defined as\footnote{A constraint of the form $y \triangleq \mathbbm{1}(x > 0)$ (where $x$ is an integer) can be expressed as the following linearized constraints: $y M \geq x$ and $y \leq x$ where $M$ is a large number.}:
\begin{equation}\label{segment_indicator}
y_{k}^{t} \triangleq \mathbbm{1} \Bigg( \sum_{ j \in \mathbb{N} }^{}{a_{j,k}^{t} {\bar{m}}_{j}^{t}} > 0 \Bigg),
\end{equation}
where ${\bar{m}}_{j}^{t} \in \{0,1\}$ signifies whether UE $j$ has successfully transmitted during time slot $t$. Based on \eqref{segment_indicator}, the normalized \textcolor{black}{semantic} throughput of each UE can be determined. Considering that $\mathbb{W}^{t} = \{ t-\mathcal{H}, \ldots, t \}$ represents a sliding time window of length $\mathcal{H}$, \textcolor{black}{$b_{k}$ is the size of segment $k$ in bits,} and ${\Omega}_{i}^{t} = 1/\big(\sum_{k, t \in \mathbb{K}, \mathbb{W}^{t}} {a_{i,k}^{t} \ {\color{black} b_{k}}}\big)$ acts as the normalization factor, the normalized \textcolor{black}{semantic} throughput of user $u_{i}$ is as follows:
\begin{equation}\label{ue_throughput}
x_{i}^{t} = {\Omega}_{i}^{t} \sum\limits_{k, t \in \mathbb{K}, \mathbb{W}^{t}} {a_{i,k}^{t} \ {\color{black} b_{k}} \ y_{k}^{t}}. \\
\end{equation}

By isolating user $u_{i}$ from other UEs in \eqref{ue_throughput}, we can break down $x_{i}^{t}$ into two components: \textit{self throughput} ($\dot{x}_{i}^{t}$) and \textit{assisted throughput} ($\ddot{x}_{i}^{t}$), defined in \eqref{ue_throughput_decomposed}. The assisted throughput for UE $i$ represents the contribution of other UEs in transmitting shared segments between them and user $u_i$ to the SBS.
\begin{align}\label{ue_throughput_decomposed}
    & {x}_{i}^{t} \ \myeq \ \dot{x}_{i}^{t} + \ddot{x}_{i}^{t} \notag \\
    & \dot{x}_{i}^{t} \ = \ {\Omega}_{i}^{t} \sum\limits_{k,t \in \mathbb{K}, \mathbb{W}^{t}} {a_{i,k}^{t} \ {\color{black} b_{k}} \  {\bar{m}}_{i}^{t}} \\
    & \ddot{x}_{i}^{t} \ = \ {\Omega}_{i}^{t} \sum\limits_{k,t \in \mathbb{K}, \mathbb{W}^{t}} { \Big( a_{i,k}^{t} \ {\color{black} b_{k}} \ (1 - {\bar{m}}_{i}^{t}) y_{k}^{t, -i} \Big) } \notag
\end{align}
Here, $y_{k}^{t, -i} = \min \big\{ 1, \sum_{ j \in \mathbb{N} \text{\textbackslash} \{ i \} }^{}{a_{j,k}^{t} {\bar{m}}_{j}^{t}} \big\}$\textcolor{black}{, indicating the transmission of segment $k$ at time $t$ by all users except user $i$; if more than one user transmits, only one transmission is counted}.
\vspace{5 pt}

\begin{table*}[!hbt]
\caption{Potential Use Cases for Our Framework.}
\centering
\small
\renewcommand{\arraystretch}{1.1}
\begin{tabularx}{\textwidth}{|X|X|X|X|}
\hline
\multicolumn{1}{|c|}{\textbf{Item}} &
  \multicolumn{1}{c|}{\textbf{Task}} & \textbf{Semantic segment} &
  \textbf{\textcolor{black}{Example construction of $\mathbb{A}$}} \\
\hline
Multi-UAV Area Coverage \cite{11142806, 10664472, mazandarani2025semantic} &  \textcolor{black}{UAVs send overlapping views of the same area.}
  & \textcolor{black}{Objects or Image regions} &
  \textcolor{black}{If two UAVs detect the same object in the overlap, both are linked to the same segment (illustrated in Fig. \ref{illustrative_example}).} \\
\hline
Holographic Presence \cite{haoyu2023xr} &  \textcolor{black}{XR users share parts of their local scenes.}
 & \textcolor{black}{Objects and spatial relations} &
 \textcolor{black}{Matching the same object or relation, e.g., table-next-to-chair, creates a shared segment.} \\ 
\hline
IoT Monitoring System \cite{kassab2020multi} &  \textcolor{black}{Sensors monitor the same physical event.}
 & \textcolor{black}{Events, e.g., high temperature} &
 \textcolor{black}{Sensors reporting the same event in the same area and time window share a segment.}  \\
\hline
Image Segmentation Reconstruction \cite{pan2023image} &  \textcolor{black}{UEs send parts of an image segmentation map.}
 & \textcolor{black}{Image masks or regions} &
 \textcolor{black}{Masks with the same class and strong overlap are mapped to the same segment.} \\ 
\hline
Cooperative Navigation \cite{abdel2024cooperative} &  \textcolor{black}{Agents share local map observations.}
 & \textcolor{black}{Grid-cell states, e.g., obstacle/free} &
 \textcolor{black}{Reports for the same grid cell and state are assigned to the same segment.} \\ 
\hline
6G Digital Twin \cite{lin20236g} & \textcolor{black}{Devices update the same digital replica.}
 & \textcolor{black}{Twin objects or states} &
 \textcolor{black}{Reports about the same component or state are mapped to the same segment.} \\ 
\hline
Collaborative GenAI \cite{du2023exploring} &  \textcolor{black}{Users generate related parts of a shared output.}
 & \textcolor{black}{Generated content blocks} &
 \textcolor{black}{Users working on the same content block (e.g, a room) share a segment.} \\ 
\hline
Satellite Earth Observation \cite{bui2024semantic} &  \textcolor{black}{Satellites report changes over Earth regions.}
& \textcolor{black}{Geo-tagged regions or events} &
\textcolor{black}{Detections with the same location, event type, and time window share a segment.} \\ 
\hline
\end{tabularx}
\label{scenarios}
\end{table*}

\subsubsection{Correlated Utilities}

Following the defined semantic correlations among users, it is evident that users' utilities are not independent, as commonly assumed in traditional networks. Specifically, by allocating more resources to user $i$ at time $t+1$, the utility of another user $j$ may also increase. Mathematically, this can be expressed as $\exists \ i, \ j \neq i, t: \ \mathfrak{R}_{i}^{t+1} > \mathfrak{R}_{i}^{t} \longrightarrow \mathcal{U}_{j}^{t+1} > \mathcal{U}_{j}^{t}$, where $\mathfrak{R}_{i}^{t}$ and $\mathcal{U}_{i}^{t}$ denote the resources allocated to and the utility of user $i$ at time $t$, respectively. This utility correlation is observed in various real-world scenarios, including the setup from our previous work on Multi-UAV Area Coverage \cite{mazandarani2025semantic}. In this setup, multiple UAVs collaboratively transmit partially overlapping observations to a central server. A transmission by each UAV may lead to a better reconstructed image and, consequently, more effective tasks like object detection for nearby UAVs.




Although shared semantic information exists among user subsets in all these scenarios, we are not restricted to situations where only a subset of users share data; users may also possess information that plays an equally important role in task completion. For instance, in a collaborative  GenAI setup \cite{du2023exploring}, users work together to create synthetic data. Suppose the responsibility of generating a certain subspace of the data is allocated to more than one user. In that case, semantic-aware decision-making will enable the selection of users based on alternative factors such as fairness or users' energy levels.

Table \ref{scenarios} summarizes additional scenarios and further details. \textcolor{black}{In general, this framework applies to any contention-based many-to-one uplink transmission with local access feedback, where a service-specific rule maps UE descriptors to common segment identities, concurrent duplicate deliveries of a segment are counted only once, and semantic overlap is non-negligible and persists during policy learning. In other words, the current framework offers no semantic-redundancy gain for independent payloads or when encryption prevents segment matching, and it is not intended for downlink or centrally scheduled access, hard per-packet guarantees, or services requiring per-UE raw-data provenance.}
\vspace{3 pt}

\subsubsection{Fairness-utilization trade-off} To ensure fairness among UEs while maximizing network utilization, the $\alpha$-fairness metric can be used as the objective function, as defined by the following formula.
\begin{align}\label{alpha_fairness}
    & \mathcal{F}_{\alpha}^{t} = 
    \begin{cases} 
        \sum\limits_{i \in \mathbb{N}}{ log(x_{i}^{t})} & \text{if } \alpha=1 \\
        (1 - \alpha)^{-1} \sum\limits_{i \in \mathbb{N}}{ {(x_{i}^{t})}^{1-\alpha}} & \text{if } \alpha \neq 1
    \end{cases}
\end{align}
When $\alpha \to 0$, optimizing \eqref{alpha_fairness} becomes equivalent to maximizing the sum of throughputs, commonly known as utilitarian fairness or sum-rate maximization. In this scenario, the objective is to maximize the total throughput, with minimal consideration for its distribution among users. Although this method is generally more efficient, it can result in significant inequality. Conversely, when $\alpha \to \infty$, optimizing \eqref{alpha_fairness} shifts to maximizing the minimum user's throughput (i.e., $\max \min_{i}{x_{i}^{t}}$), known as max-min fairness. Max-min fairness ensures that throughput allocations are adjusted to enhance the allocation received by the user with the least resources, thereby achieving the highest level of equality among users, albeit potentially at the expense of overall efficiency.

\textcolor{black}{One should note that even within a single application, UE contributions are only partly redundant, as each UE also holds segments that no other UE can deliver, and each access attempt draws transmission and encoder-inference energy from a battery-limited device; $\alpha$ therefore controls whether access prioritizes aggregate semantic throughput ($\alpha \to 0$) or limits the persistent starvation of individual UEs and the energy concentration it implies. Furthermore, a single application does not imply a single utility holder, as the UEs of one service may belong to distinct stakeholders whose individual utilities remain separately meaningful, such as the portion of a holographic scene that is private to one participant.}
\vspace{3 pt}

\subsubsection{Energy Awareness}

To maximize network utilization while minimizing energy consumption, we \textcolor{black}{adopt a linearized energy-efficiency utility and define it for} user $u_{i}$ as the difference between its throughput at time slot $t$ (i.e., ${x}_{i}^{t}$) and a coefficient $\zeta$ of its consumed energy. The consumed energy includes both packet transmissions, which are a function of its own throughput, and semantic encoder inferences, which are a function of the number of decision makings (attempts to transmit on any channel) in $\mathbb{W}^{t}$, denoted with ${D}_{i}^{t}$.
\begin{equation} \label{energy_utility}
    {e}_{i}^{t} =  {{x}_{i}^{t}} - \zeta \big( { g( \dot{x}_{i}^{t} ) + h( {D}_{i}^{t}) } \big)
\end{equation}
Assuming a simple premise, we posit in this paper that $g(\dot{x}_{i}^{t}) = {\sigma}_{tr} \dot{x}_{i}^{t}$ and $h({D}_{i}^{t}) = {\sigma}_{inf} {D}_{i}^{t} + {\sigma}_{init}$, where ${\sigma}_{tr}$ denotes the energy consumption for transmission during a single time slot, ${\sigma}_{inf}$ represents the energy required for the semantic extraction of one packet (i.e., inference on the encoder), and ${\sigma}_{init}$ denotes the energy needed to initialize the semantic encoders.

In a bit-oriented network, ${\sigma}_{inf}$ and ${\sigma}_{init}$ lack relevance. However, in our energy-aware semantic-oriented setting, the relative importance of ${\sigma}_{inf}$ and ${\sigma}_{init}$ compared to ${\sigma}_{tr}$ significantly impacts users' utilities and their behavior. Consequently, the energy efficiency of the system is determined by the sum of the energy efficiencies of the UEs. In the following formula, $d_{1} = \frac{{\sigma}_{inf}}{{\sigma}_{tr}}$ and $d_{2} = \frac{{\sigma}_{init}}{{\sigma}_{tr}}$.
\begin{equation} \label{energy_utility_normalized}
    \mathcal{E}_{\zeta, d_{1}, d_{2}}^{t} =  \sum\limits_{i \in \mathbb{N}} \Big( {{{x}_{i}^{t}} - { \zeta \big( \dot{x}_{i}^{t} + ( d_{1}  {D}_{i}^{t} ) + d_{2} } \big) \Big) }
\end{equation}
As $\zeta \to 0$, optimizing \eqref{energy_utility_normalized} simplifies to a sum-rate maximization problem (akin to the scenario in \eqref{alpha_fairness} as $\alpha \to 0$). By increasing $\zeta$, the consideration of energy consumption becomes more important in the decision-making, ultimately leading to inactive users avoiding negative objective values (since the maximum of a negative value is zero). Therefore, $\zeta$ should be thoughtfully adjusted based on the system characteristics and goals. Another determining parameter is $d_{1}$, and its value influences the average packet size that users transmit (as investigated in section \ref{opt_sol}.). It is worth noting that in the special case where $d_{1}$ and $d_{2}$ equals zero (i.e., $\mathcal{E}_{\zeta, d_{1}, d_{2}}^{t} =  \sum\limits_{i \in \mathbb{N}} \big( {{{x}_{i}^{t}} - { \zeta \dot{x}_{i}^{t} }  \big) }$), complexity persists since users' energy consumption is a function of self throughputs.

{
\color{black}

\subsubsection{Practical Challenges of Semantic Orchestration}

Before formulating the problem, it is worth mentioning the practical challenges that arise in defining semantic segments and association matrices. The KB-MANO framework \cite{our_mag_paper} can be utilized as the semantic orchestration entity to address these challenges. Within this framework, the construction of the association matrix begins with a service-specific definition of a semantic segment. The appropriate granularity depends on the downstream task. In collaborative object detection, a segment may correspond either to an object class or to the presence of an object in a region. For scene reconstruction tasks, however, more descriptive semantic segments (e.g., object embeddings within the scene) improve the fidelity of the inferred correlations but also increase the number of candidate segments, the matching complexity, and the signaling overhead. Hence, the segment vocabulary $\mathbb{K}$ and its granularity are service-level design parameters rather than universal quantities.

Another challenge is that the base station, which is responsible for reconstructing users' data or performing downstream tasks on it (e.g., object detection or visual question answering), as well as rewarding users based on their assisting data, does not have access to users' data in advance to construct the association matrix. To address this, the base station can collect semantic metadata from the UEs (e.g., a segment identifier, a quantized embedding, a hash, or a confidence score, rather than the raw observation), assuming that the overhead of transmitting them is negligible. The base station then forwards this metadata to the KB MANO framework, which resolves descriptor identities across users and constructs the estimated association matrix. In slowly varying settings, the matrix can be propagated from previous time slots using object tracking, mobility prediction, or temporal smoothing, with event-triggered recomputation when the observed confidence or application-level QoS/QoE falls below a prescribed threshold. The present optimization uses the estimated association matrix as a deterministic input and does not explicitly account for descriptor-signaling overhead or matrix-estimation errors.

}

\subsection{Problem Formulation}

In this subsection, we present a Mixed-Integer Non-Linear Programming (MINLP) formulation to define two centralized optimization problems. \textcolor{black}{This formulation serves as a centralized collision-free benchmark rather than a distributed MAC protocol. In particular, the optimizer is assumed to have system-wide information over the optimization horizon and to schedule packet starts, packet lengths, and channels. Collisions are therefore excluded by construction. In contrast, collisions may occur in the distributed PRISM protocol introduced in Section~\ref{S_SES}, where UEs independently select their actions from local observations.}

The subsequent subsection will further elaborate on the problems' variables and constraints, as well as the objective function. Notably, the variable packet-length formulation is an enhancement of our previous work \cite{TMLCN_2024} to a multi-agent setup.

\subsubsection{Decision and Support Variables}

The primary integer decision variable in our problems is $\mathbb{\mathfrak{R}} = [{r}_{i, c}^{t}]_{\mathcal{N} \times \mathcal{C} \times \mathcal{T}}$, where each element represents the size of the packet that user $u_{i}$ begins to transmit on channel $c$ at the start of time slot $t$. Additionally, $\mathbb{Z} = [{z}_{i, c}^{t}]_{\mathcal{N} \times \mathcal{C} \times \mathcal{T}}$ denotes an auxiliary integer variable indicating the remaining time slots for the ongoing transmission. The variable $\mathbb{M} = [{m}_{i, c}^{t}]_{\mathcal{N} \times \mathcal{C} \times \mathcal{T}}$ serves as another auxiliary variable, set to one whenever ${z}_{i, c}^{t}$ is positive. 
\textcolor{black}{More specifically, ${m}_{i,c}^{t}$ is a channel-occupancy indicator for UE $i$ on channel $c$ at time slot $t$. It is not an independently selected scheduling action; rather, it is determined by the remaining-duration variable ${z}_{i,c}^{t}$. If ${z}_{i,c}^{t}=0$, then UE $i$ is not occupying channel $c$ in slot $t$, and therefore ${m}_{i,c}^{t}=0$. If ${z}_{i,c}^{t}>0$, then UE $i$ is still transmitting a packet on channel $c$, and therefore ${m}_{i,c}^{t}=1$. For example, if UE $i$ starts a packet of length 3 on channel $c$ at time slot $t$, then the corresponding occupancy indicator remains 1 for the next 3 occupied slots. This indicator is later used in the channel-occupancy constraints to prevent multiple UEs from occupying the same channel in the centralized benchmark.}
The following constraints transform the equation ${m}_{i, c}^{t} = \mathbbm{1}({z}_{i, c}^{t} > 0)$ into a linear form:
\begin{align}\label{z_m_constraints}
    &{m}_{i, c}^{t} \leq {z}_{i, c}^{t}, \\
    &{m}_{i, c}^{t} R_{max} \geq {z}_{i, c}^{t}, \notag
\end{align}
where $R_{max}$ is the maximum allowed packet size.

\subsubsection{Temporal Interdependency Constraints}

Now, we need to establish interdependency over time. First, we must ensure that $z$ decreases by one with each passing time slot for ongoing transmissions. This requirement is enforced by the following constraint.
\begin{align}\label{z_constraints}
{z}_{i, c}^{t} = {z}_{i, c}^{t - 1} - 1 \quad \text{if } \ \big( ( {m}_{i, c}^{t} = 1 ) \ \land \ ( {d}_{i, c}^{t} = 0 ) \big)
\end{align}
This constraint can be expressed in its linear equivalent forms as follows:
\begin{align}\label{z_constraints_linear}
    &{z}_{i, c}^{t} \leq ( {z}_{i, c}^{t - 1} - 1 ) + \left( R_{max} \big( (1 - {m}_{i, c}^{t} ) + {d}_{i, c}^{t} \big) \right), \\
    &( {z}_{i, c}^{t - 1} - 1 ) \leq {z}_{i, c}^{t} + \left( R_{max} \big( (1 - {m}_{i, c}^{t} ) + {d}_{i, c}^{t} \big) \right), \notag
\end{align}
where $\mathbb{D} = [{d}_{i, c}^{t}]_{\mathcal{N} \times \mathcal{C} \times \mathcal{T}}$ is a binary variable used to activate decision making (i.e., starting packet transmission) over specific time slots.

Second, it must be ensured that a new transmission is not initiated until the ongoing one is completed. To achieve this, we establish a set of constraints based on the premise that whenever user $u_{i}$ initiates the transmission of a packet with length $k$ at time slot $t$, the decision-making variables for the subsequent $k-1$ time slots must be zero (i.e., ${{d}_{i, c}^{\tau}} = 0 \ \forall \tau \in [t+1, t+k-1]$):
\begin{align}\label{r_d_constraints}
    {r}_{i, c}^{t} = k \to  \sum_{\tau=t+1}^{t+k-1}{{d}_{i, c}^{\tau}} = 0 . \quad \forall k \in [1, R_{max}]
\end{align}
This can be expressed in its linear equivalent form as follows (where $K_{1}$ and $K_{2}$ are auxiliary binary variables):
\begin{align}\label{r_d_constraints_linear}
    &( {r}_{i, c}^{t} - k ) + 1 \leq R_{max} \ K_{1} \\
    &( k - {r}_{i, c}^{t} ) + 1 \leq R_{max} \ K_{2} \\ 
    &\sum_{\tau=t+1}^{t+k-1}{{d}_{i, c}^{\tau}} \leq R_{max} (2 - ( K_{1} + K_{2}))
\end{align}

\subsubsection{Packet Length Constraints}

\begin{figure}[t!]\centering
\includegraphics[width=3.0in]{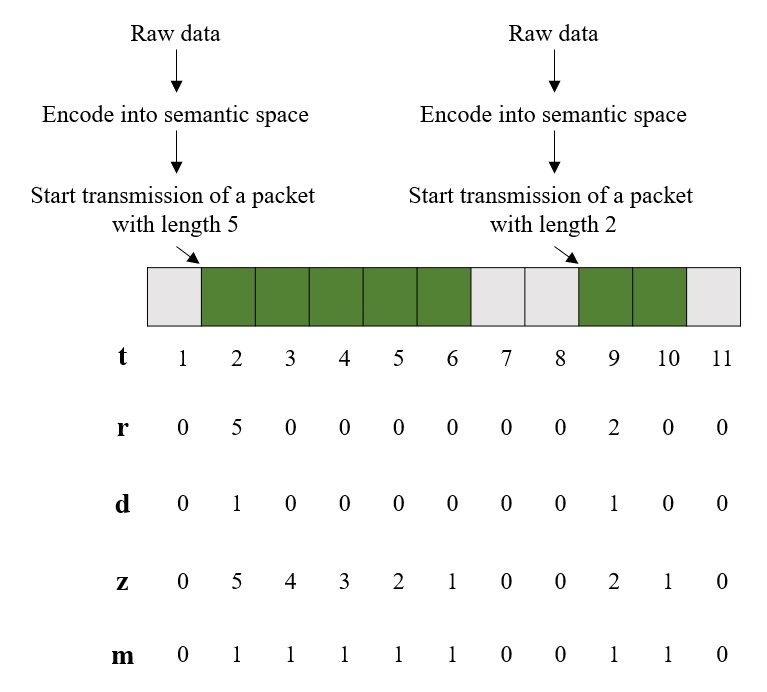}
    \vspace{-5pt}
    \caption{A scenario in which a UE initiates the transmission of packets with lengths 5 and 2 at time slots $t = 2$ and $t = 9$, respectively. The values of the decision variable $r$, along with the support variables $d$, $z$, and $m$, can be compared accordingly.}
    \vspace{0pt}
    \label{variables}
\end{figure}

Next, we establish the relationship between the packet length variable $r$ and the support variables $z$ and $d$ using \eqref{r_constraints}, where $r$ is equal to $z$ when transmission starts:
\begin{align}\label{r_constraints}
{r}_{i, c}^{t} = \begin{cases} {z}_{i, c}^{t} & \text{if } \ {d}_{i, c}^{t} = 1 \\
0 & \text{if } \ {d}_{i, c}^{t} = 0 \end{cases}
\end{align}
We then transform \eqref{r_constraints} into a linear form, similar to previous constraints:
\begin{align}\label{r_constraints_linear}
    &{r}_{i, c}^{t} \leq {z}_{i, c}^{t}  + \left( R_{max} (1 - {d}_{i, c}^{t} ) \right) \\
    & {z}_{i, c}^{t} \leq {r}_{i, c}^{t}  + \left( R_{max} (1 - {d}_{i, c}^{t} ) \right) \notag
    \\
    & {r}_{i, c}^{t} \leq R_{max} {d}_{i, c}^{t} \notag
\end{align}
A simple example, as illustrated in Fig. \ref{variables}, demonstrates the values of $r$ and the other auxiliary variables. It is important to note that in a single time-slot scenario, the variables $r$, $z$, $m$, and $d$ are equal and binary.

\begingroup
\begin{table}[t!]
\caption{Notations of Symbols Used in the Problem Definitions.}
\vspace{-1em}
\setlength\tabcolsep{2.0pt}
\setlength\extrarowheight{-3pt}
\renewcommand{\arraystretch}{2.0}
\begin{center}
\small
\begin{tabular}{|c|l|}
\hline

\textbf{Symbol} & \multicolumn{1}{c|}{\textbf{Description}} \\ \hline

$u_{i}$ & UE with index $i \in \mathbb{N} = \{1, \ldots, \mathcal{N}\}$ \\ \hline

$k$ & Semantic entity (or segment) $k \in \mathbb{K} = \{1, \ldots, \mathcal{K} \}$ \\ \hline

$y_{k}^{t}$ & Segment transmission indicator for segment $k$ at time slot $t$ \\ \hline

$a_{i,k}^{t}$ & Indicator for association of $u_{i}$ with segment $k$ at time slot $t$ \\ \hline

${\bar{m}}_{i}^{t}$ & \makecell[l]{Indicator for successful transmission of $u_{i}$ \\ during time slot $t$ over all channels (i.e., $\sum_{c \in \mathbb{C}}^{}{{m}_{i,c}^{t}}$)} \\ \hline

$\mathbb{W}^{t}$ & Sliding time window of length $\mathcal{H}$, i.e.,  $ \{ t-\mathcal{H}, \ldots, t \}$ \\ \hline

${x}_{i}^{t}$ & Actual throughput of $u_{i}$ in $\mathbb{W}^{t}$ \\ \hline

${\Omega}_{i}^{t}$ & Normalization factor (i.e., $1/\big(\sum_{k, t \in \mathbb{K}, \mathbb{W}^{t}} {a_{i,k}^{t}}\big)$) \\ \hline

$\dot{x}_{i}^{t}$ & Self throughput of $u_{i}$ in $\mathbb{W}^{t}$ \\ \hline

$\ddot{x}_{i}^{t}$ & Assisted throughput of $u_{i}$ in $\mathbb{W}^{t}$ \\ \hline

$\mathcal{F}_{\alpha}^{t}$ & $\alpha$-fairness
metric in $\mathbb{W}^{t}$, as defined in \eqref{alpha_fairness} \\ \hline

${e}_{i}^{t}$ & Energy efficiency of $u_{i}$ in $\mathbb{W}^{t}$ \\ \hline

${D}_{i}^{t}$ &  \makecell[l]{Number of decision makings \\ (attempts to transmit on any channel) in $\mathbb{W}^{t}$ } \\ \hline

$\mathcal{E}_{\zeta, d_{1}, d_{2}}^{t}$ & Energy efficiency of the system in $\mathbb{W}^{t}$, as defined in \eqref{energy_utility_normalized} \\ \hline

${m}_{i,c}^{t}$ & \makecell[l]{An indicator for transmission of UE $i$ \\ on channel $c$ in time slot $t$} \\ \hline

${z}_{i, c}^{t}$ & \makecell[l]{Number of time slots remaining \\ until the current transmission of $u_{i}$ on channel $c$ ends} \\ \hline

${R}_{max}$ & maximum allowed packet size \\ \hline

${r}_{i, c}^{t}$ & \makecell[l]{Size of the packet $u_{i}$ starts transmitting \\ on channel $c$ at the beginning of time slot $t$} \\ \hline

${d}_{i,c}^{t}$ & \makecell[l]{An indicator for decision-making of UE $i$ \\ on channel $c$ in time slot $t$} \\ \hline

\textcolor{black}{${Q}_{c}^{t}$} & \makecell[l]{\textcolor{black}{Quality indicator for channel $c$ in time slot $t$}} \\ \hline

\end{tabular}
\label{tab_symbols}
\end{center}
\end{table}
\endgroup

\subsubsection{Physical Constraints}

\textcolor{black}{Final constraints, \eqref{max_channel_constraints} and \eqref{max_user_constraints}, ensure that each user transmits on at most one channel per time slot, and that no more than one user transmits simultaneously on the same channel that satisfies the required quality condition, as indicated by ${Q}_{c}^{t}$.}
\begin{equation} \label{max_channel_constraints}
    \sum_{c \in \mathbb{C}} {m}_{i, c}^{t} \leq 1 \quad \forall i \in \mathbb{N}, \forall t \in [0, \mathcal{T}]
\end{equation}
\begin{equation} \label{max_user_constraints}
    \sum_{i \in \mathbb{N}} {m}_{i, c}^{t} \leq \textcolor{black}{{Q}_{c}^{t}} \quad \forall c \in \mathbb{C}, \forall t \in [0, \mathcal{T}]
\end{equation}

\subsubsection{Problems}

Finally, we address two non-linear problems to maximize $\alpha$-fairness and \textcolor{black}{linearized system energy efficiency utility} (presented in $P_{1}$ and $P_{2}$, respectively). The non-linearity of both problems arises from the semantic interdependence of packet transmissions. In particular, the definition of $\mathcal{F}_{\alpha}^{t}$ remains non-linear \textcolor{black}{except for $\alpha = 0$}, necessitating piecewise linear approximation techniques \cite{eriksson2004piecewise}.
\begin{align} \label{problems}
    &\max_{R} \ \mathcal{F}_{\alpha}^{T}  \quad \quad \ \ \mbox{acc. \eqref{alpha_fairness}}  \tag{$P_{1}$} \\
    &\mbox{s.t.
    \eqref{z_m_constraints}, \eqref{z_constraints_linear}, \eqref{r_d_constraints}, \eqref{r_constraints_linear}, \eqref{max_channel_constraints}, \eqref{max_user_constraints}} \notag \\ \notag \\
    &\max_{R} \ \mathcal{E}_{\zeta, d_{1}, d_{2}}^{T} \quad
    \mbox{acc. \eqref{energy_utility_normalized}} \tag{$P_{2}$} \\
    &\mbox{s.t.
    \eqref{z_m_constraints}, \eqref{z_constraints_linear}, \eqref{r_d_constraints}, \eqref{r_constraints_linear}, \eqref{max_channel_constraints}, \eqref{max_user_constraints}} \notag
\end{align}

\subsection{Optimal Solutions} \label{opt_sol}

\begin{figure}[t!]\centering
\includegraphics[width=3.5in]{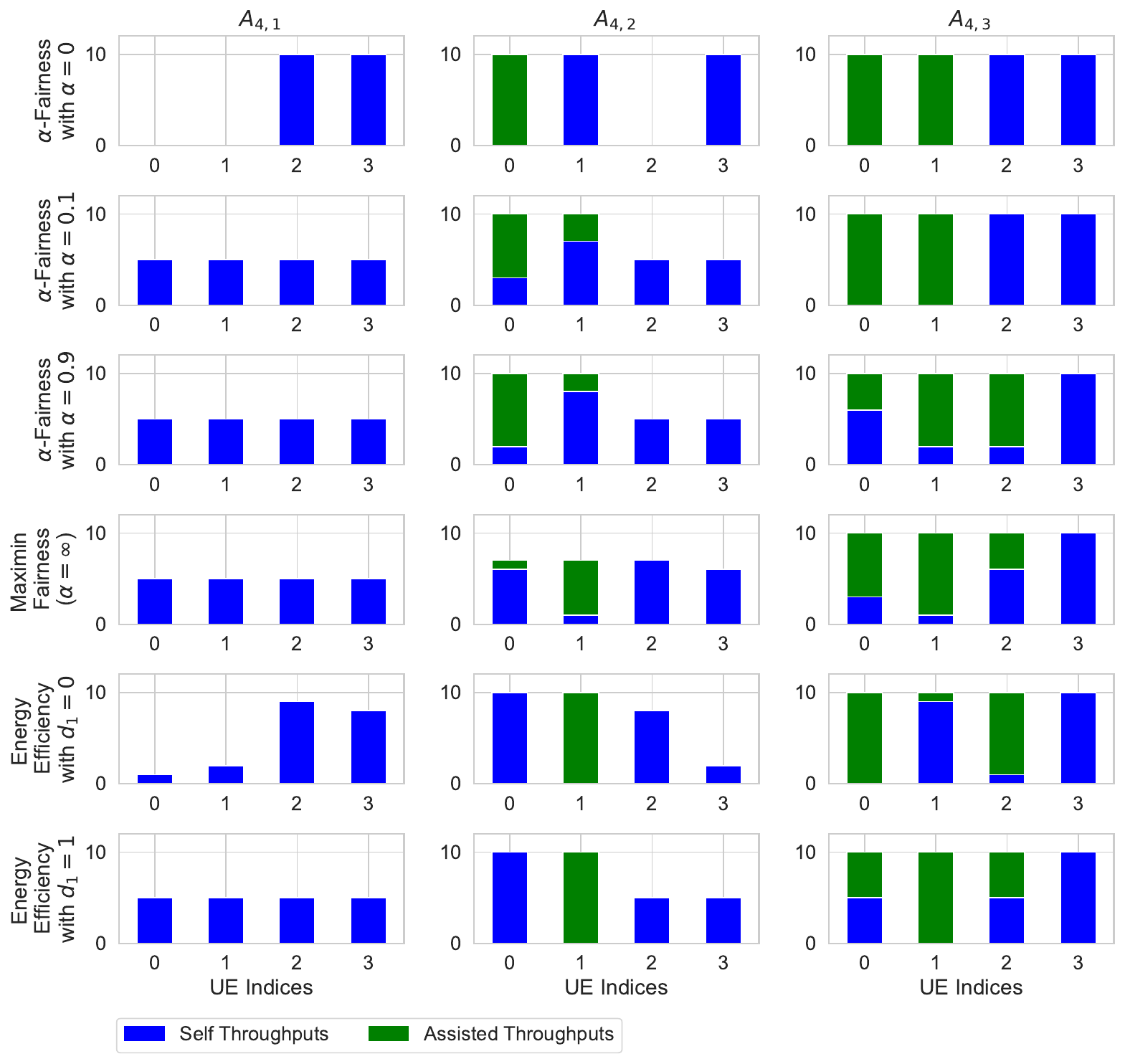}
    \vspace{-5pt}
    \caption{Optimal self and assisted UE throughputs for different objective functions and association matrices. Note that these optimal points are not necessarily unique.}
    \vspace{-15pt}
    \label{fig_optimal}
\end{figure}

Despite our efforts to linearize the constraints of problems \( P_{1} \) and \( P_{2} \) and employ a linear objective function for energy efficiency, solving them remains intractable due to the presence of multiple integer and binary variables. The solution space for the integer variables $ r_{i, c}^{t}$ and $z_{i, c}^{t}$ is $(R_{max} + 1)^{\mathcal{N} \cdot \mathcal{C} \cdot \mathcal{T}}$, and for the binary variables $m_{i, c}^{t}$ and $d_{i, c}^{t}$ is $2^{\mathcal{N} \cdot \mathcal{C} \cdot \mathcal{T}}$. Considering all these variables, their solution space sizes must be multiplied together to calculate the solution space of the problems, rendering them impractical to address large numbers of users, channels, maximum packet sizes, or network lifetime. However, to gain insights into problems, we solve them by utilizing the \textcolor{black}{Gurobi optimizer\footnote{Gurobi Optimizer, {https://www.gurobi.com}.}} for a modest setting with $\mathcal{N} = 4$, $\mathcal{C} = 2$, $R_{\max} = 5$, $\mathcal{T} = 10$, and various association matrices $A_{\mathcal{N}, q}$, where the first $q$ UEs of $\mathcal{N}$ UEs share their segments ($A_{\mathcal{N}, 1}$ and $A_{\mathcal{N}, \mathcal{N}}$ represent the scenarios of no semantic sharing and full semantic sharing, respectively). The results of these simulations are interpreted and visualized in Fig. \ref{fig_optimal}.

With increasing $q$ in $A_{q}$, both objective functions $\mathcal{F}_{\alpha}^{t}$ and $\mathcal{E}_{\zeta, d_{1}, d_{2}}^{t}$ improve because users assist each other in their communications. However, this positive impact varies depending on the objective functions and their parameters. Our observations revealed that metrics focusing less on fairness have a greater potential for improvement since allocating additional resources to users who share segments adds greater value. Precisely, considering the standard deviation among optimal UE throughputs as a measure of \textit{un}fairness\footnote{For example, in $\alpha$-fairness with $\alpha\to\infty$, unfairness is zero since all throughputs are equal.}, the objective improvement ratio of $A_{4, 3}$ over $A_{4, 1}$ has a positive correlation equal to $0.58$ with unfairness\footnote{As a side note, semantic awareness can potentially lead to unfairness as economic advantages tend to favor users with co-related information.}.

Finally, we provide some intuitions regarding utilization-sustainability trade-offs. In $\alpha$-fairness maximization, higher $\alpha$ leads to more equal throughputs among users, hence higher societal sustainability. In energy efficiency maximization, when transcoding costs are considered in the objective function ($d_{1} = 1$), it results in larger packets. Numerically, the average packet size for energy efficiency with $d_{1} = 1$ is $5$ while for $d_{1} = 0$ equals $2.5$.

%% file: sections/4_PRISM.tex
\section{PRISM}\label{S_SES}

{
\color{black}
The optimization problems \(P_1\) and \(P_2\) characterize the desired semantic-aware allocation, but they are not suitable for online MAC operation. First, their search spaces grow exponentially with the number of UEs, channels, time slots, and admissible packet lengths, as discussed in Section~\ref{opt_sol}. Solving them also requires centralized, system-wide information over the optimization horizon. In contrast, a UE must make each access decision promptly from local observations and without knowing the concurrent decisions of the other UEs. Conventional bit-oriented MAC schemes are likewise insufficient because they treat successful packets as independent contributions: they cannot determine whether transmitting a packet delivers new semantic segments or merely repeats segments already supplied by another UE. Moreover, a decision must jointly select a channel and a packet length. Longer packets may improve utilization and amortize semantic-encoding costs, but they occupy a channel for multiple time slots and expose the UE to a longer collision interval. These coupled semantic, temporal, fairness, and energy effects make a fixed or myopic access rule difficult to design.

The resulting decision process is naturally modeled as a Macro-Action Decentralized Partially Observable Markov Decision Process (MacDec-POMDP). A MacDec-POMDP extends a Dec-POMDP by allowing actions with variable durations, such as the multi-slot packet transmissions considered here \cite{xiao2020macro}. Partial observability arises because each UE observes only the channel that it senses or uses, while the success or collision outcome of a multi-slot transmission is revealed only after that transmission terminates. A learning-based solution is therefore needed to infer effective access policies from interaction histories rather than repeatedly solving the full optimization problem. We adopt MADRL because it supports decentralized decisions under local observations while allowing the agents to learn coordinated behavior from a common objective during centralized training. Accordingly, we propose PRISM (\underline{P}rotocol for \underline{R}edundancy \underline{I}dentification in \underline{S}emantic \underline{M}ultiple-access), in which one DRL agent controls the MAC decisions of each UE. PRISM explicitly exposes self and assisted semantic transmissions to the agents, represents variable-length transmissions as macro-actions, and shapes the common reward according to either the fairness-utilization or energy-efficiency objective. Thus, it preserves the objectives of \(P_1\) and \(P_2\) while providing an online, decentralized execution mechanism.
}


In this context, agent \( i \)'s macro-action \( a_{i} \) is defined as a tuple \(\left\langle I_{a_{i}}, \pi_{a_{i}}, \beta_{a_{i}} \right\rangle\), representing the initiation set, macro-action low-level policy, and termination set, respectively. In our problems, a new decision on channel sensing or packet transmission (denoted by the variable \( r \) in the optimization problems) is made immediately after the preceding macro-action is completed. During the multi-time slot transmission process, the remaining portion of the packet is incrementally reduced until the transmission is complete (i.e., when the variable \( z \) reaches zero). For simplicity, we will refer to macro-actions as actions in the remainder of this paper. Furthermore, we define the set of active UEs at time slot \( t \), denoted as \(\mathbb{N}^{t}_{+}\), as the subset of UEs that have completed their actions and are ready for new transmissions. Formally, this is represented as: \(\mathbb{N}^{t}_{+} = \{ i \in \mathbb{N} \ | \ \beta_{a_{i}} = 1 \}\). Non-active UEs are excluded from the joint action, state space, and reward calculations, as suggested by Marchesini \textit{et al.}~\cite{marchesini2023value}.

\textcolor{black}{Traffic is assumed to be saturated, in the sense that semantic content is always available at every UE, but it is generated on demand rather than queued: UE $i$ invokes its semantic encoder at the instant it initiates a transmission action, and the resulting descriptor supersedes any older unsent one, so no backlog is maintained. A sensing action neither invokes the encoder nor alters the buffered descriptor. All segments associated with the current descriptor are carried in a single packet, as assumed in \eqref{segment_indicator}, and a collided packet is not retransmitted. Finally, a UE neither observes nor assumes a neighbor's current payload; semantic overlap is resolved at the SBS after reception.} In the following sections, we will elaborate on the action and state spaces, reward mechanisms, and the architectural framework of the learning process.

\subsubsection{Action Space}

The action space for active UEs at time slot $t$ is specified in \eqref{action_space}. In this set, the action of $u_{i}$ can take one of two forms: $a_{i}: (0, c_{i})$, which indicates sensing channel $c_{i}$, or $a_{i}: (r_{i} > 0, c_{i})$, which denotes the transmission of a packet with length $r_{i}$ on channel $c_{i}$. 
\begin{equation}\label{action_space}
\small
\boldsymbol{\mathcal{A}^{t}} = \Bigg\{ \bigg\{ \boldsymbol{a}_{i}: (r_{i}, c_{i}) \big| 
\begin{array}{c}
r_{i} \in \{0, ..., {R}_{max}\}, \\ 
c_{i} \in \{1, ..., \mathcal{C} \} 
\end{array} 
\bigg\} \bigg| \ i \in \mathbb{N}^{t}_{+} \Bigg\}
\end{equation}

\textcolor{black}{Here, $a_i:(0,c_i)$ is an optional one-slot carrier-sensing action and is not mandatory before transmission. It reports only whether $c_i$ is busy or idle and reveals neither the transmitter nor its payload; a busy result therefore does not imply that a semantically related UE is transmitting. Moreover, the channels are semantically agnostic and semantic relationships are represented by $\mathbb{A}$, not by carrier assignment.}

\subsubsection{State Space}

The state space consists of five distinct components. To provide UEs with real-time fairness information, we incorporate the Delay to Last Successful Transmission (D2LT) metric \cite{guo2022multi} represented as $v_{i}$. This metric signifies the number of time slots elapsed since the last successful transmission by $u_{i}$. Specifically, we include the normalized D2LT (i.e., $ \bar{v}_{i} = v_{i}/ \sum_{j \in \mathbb{N}}{v_{j}}$) in the state space of UEs. The next component relates to the UE's observation. When sensing channel $c$, the observation set for each UE is denoted as $\dot{o}_{i}$ = \{\textit{B: Busy, I: Idle}\}, while in the case of packet transmission, it is $\dot{o}_{i}$ = \{\textit{S: Success, C: Collision}\}. Furthermore, we introduce two other state elements called \textit{self transmission} and \textit{assisted transmission} denoted as $\ddot{o}_{i} \in \mathbb{R}_{\geq 0}$ and $\dddot{o}_{i} \in \mathbb{R}_{\geq 0}$, respectively. These variables are defined in the next section.

Given that the final element within the state space pertains to UE actions, the state of each UE comprises the most recent $\mathcal{H}_{S}$ (D2LT, observation, self and assisted transmissions, action) tuples. Additionally, the state includes a vector \({\Psi}_{i}^{t} \) that averages the observation tuples over a longer period, providing a broader context for the agent\footnote{In practice, we utilize the one-hot vector representation of observations and actions as it is more compatible with learning algorithms.}. Thus, the global system state at time slot $t$ is defined as follows:
\begin{align}\label{state_space}
& \bar{\mathbf{H}}_{L, i}^{t} = \{t - \tau | \forall 1 \leq \tau: \; (i \in \mathbb{N}^{t - \tau}_{+}) \land (|\bar{\mathbf{H}}_{L, i}^{t}| = \mathcal{H}_{L}) \} \notag \\
& \bar{\mathbf{H}}_{S, i}^{t} = \{t - \tau | \forall 1 \leq \tau: \; (i \in \mathbb{N}^{t - \tau}_{+}) \land (|\bar{\mathbf{H}}_{S, i}^{t}| = \mathcal{H}_{S}) \} \notag \\
& O_{i}^{t} = (\bar{v}_{i}^{t}, \dot{o}_{i}^{t}, {\ddot{o}_{i}}^{t}, {\dddot{o}_{i}}^{t}, {a}_{i}^{t}) \notag \\
& {\Psi}_{i}^{t} = \text{AVG}\Big\{ O_{i}^{q} | q \in \bar{\mathbf{H}}_{L, i}^{t} \Big\} \notag \\
& \boldsymbol{S}^{t} = { \bigg\{ {s}_{i}^{t} : \Big\{ O_{i}^{h} | h \in \bar{\mathbf{H}}_{S, i}^{t} \Big\} \cup {\Psi}_{i}^{t} \ \bigg| \ i \in \mathbb{N}^{t}_{+} \bigg\} }
\end{align}
In the above equation, AVG denotes the average operator, \(\bar{\mathbf{H}}_{L, i}^{t}\) represents the long-term history, encompassing \(\mathcal{H}_{L}\) time steps prior to \(t\) during which UE \(i\) decided to transmit or sense, and \(\bar{\mathbf{H}}_{S, i}^{t}\) defines a short-term history with the same definition, encompassing \(\mathcal{H}_{S}\) time steps.

\subsubsection{Self and Assisted Transmissions}
At each time slot, the \textit{self transmission} of $u_{i}$ (denoted with $\ddot{o}_{i}^{t}$) determines how much $u_{i}$ is involved in the transmission of its related segments. Similarly, the assisted transmission of $u_{i}$ (denoted with $\dddot{o}_{i}^{t}$) is the normalized summation of its segments transmitted by other UEs. In practice, calculating these metrics is not straightforward, since UEs with shared segments may not be synchronous. We first assume that the SBS only decodes a segment from the largest packets received containing that segment. The reason is that the information of shorter packets is embedded in the largest packet. Furthermore, by defining D2LT metrics for segments (similar to users' D2LTs with the same interpretation, denoted with $\delta_{k}^{t}$), The SBS calculates $\nu_{k}^{t}$ (Eq. \eqref{assisted_transmission}), which represents the portion of recently successfully received packets for segment $k$ that do not intersect with previously received packets. This value indicates the contribution of users who have successfully completed their transmission for this segment. From this, self and assisted transmissions can be determined. In the following equations, $\hat{r}_{i}^{t}$ signifies the size of the packet sent by $u_{i}$ successfully received by SBS at time $t$.
\begin{align}\label{assisted_transmission}
\nu_{k}^{t} &= \min{ { \Big\{ \delta_{k}^{t}, \max_{j}{ \{ \hat{r}_{j}^{t} \} } } \Big\} } \\
\ddot{o}_{i}^{t} &= \min{ \Big\{ \frac{{\sum_{k \in \mathbb{K}}^{}{ ( a_{i,k}^{t} \nu_{k}^{t} ) }}}{{\sum_{k \in \mathbb{K}}^{}{a_{i,k}^{t}}}}, \ \hat{r}_{i}^{t} \Big\} } \\
\dddot{o}_{i}^{t} &= \frac{{\sum_{k \in \mathbb{K}}^{}{ ( a_{i,k}^{t} \nu_{k}^{t} ) }}}{{\sum_{k \in \mathbb{K}}^{}{a_{i,k}^{t}}}} - \ddot{o}_{i}^{t}
\end{align}

As a simple illustrative example, suppose that $u_{1}$ and $u_{2}$ are associated with a shared segment $k$ and successfully finish transmission of their packets with sizes $4$ and $3$ at time slot $t$ (clearly on different channels), and the segment D2LT is $2$, indicating that $2$ time slots have elapsed since the SBS received segment $k$. In this case, $\nu_{k}^{t} = \min \big\{ 2, \max \{4, 3\}\big\} = 2$, indicating that only 2 time slots of the received packets from $u_1$ and $u_2$ uniquely contributed by these UEs to deliver segment \(k\) to the SBS. Therefore, $\ddot{o}_{1}^{t} = \ddot{o}_{2}^{t} = 2 $ and $\dddot{o}_{1}^{t} = \dddot{o}_{2}^{t} = 0 $. Now assume that $u_{2}$ was not successful in its transmission. In this case, $\ddot{o}_{1}^{t} = \dddot{o}_{1}^{t} = 2 $ and $\ddot{o}_{2}^{t} = \dddot{o}_{2}^{t} = 0 $. Finally, assume that the D2LT is 10, meaning that segment \(k\) has not been seen by the SBS recently. In this case, $\nu_{k}^{t} = \min \big\{ 10, \max \{ 4, 3 \} \big\} = 4$, indicating that all packets can be considered as unique contributions. Therefore, \(\ddot{o}_{1}^{t} = 4\) and \(\ddot{o}_{2}^{t} = 3\), resulting in \(\dddot{o}_{1}^{t} = 0\) and \(\dddot{o}_{2}^{t} = 1\). 

\textcolor{black}{It should be noted that a UE is not given its neighbors' data, actions, or association rows. The sensing outcome $\dot{o}_i^t\in\{B,I\}$ is obtained locally. After each completed macro-action, the SBS uses $\mathbb{A}$ and $\delta_k^t$ to compute $\ddot{o}_i^t$ and $\dddot{o}_i^t$ and returns them with $\bar{v}_i^t$; after a transmission, it also returns $\dot{o}_i^t\in\{S,C\}$. The policy therefore estimates semantic contribution from the history in \eqref{state_space}, rather than receiving it in advance. This feedback is lightweight and downlink-only: one short control message per completed macro-action rather than per time slot, which terminates the macro-action where an acknowledgment would and can thus be piggybacked on it \cite{naparstek2018deep, guo2022multi, miuccio2024learning}, without consuming the $\mathcal{C}$ contended uplink channels. We assume this control feedback is error-free and available before the next decision; therefore delay, loss, and quantization of feedback values are outside the present model.}

\subsubsection{Reward}

Since we consider a cooperative scenario in which all UEs aim to maximize the global network objective function, the system reward is defined as the average of user rewards, as outlined in \eqref{reward}. In this calculation, \textcolor{black}{the reward of user $i$ at time step $t$ is determined by the selected action and the resulting system state. For a successful action ($\dot{o}_i^t=S$), the user receives the normalized utility $r_i^t/R_{\max}$, reduced by a strategy-dependent penalty. Under $\mathcal{F}_{\alpha}^{t}$, the term $\alpha\max_{j}\{\bar{v}_{j}^{t}\}$ penalizes the maximum constraint violation. Under $\mathcal{E}_{\zeta,d_1,d_2}^{t}$, the penalty $\zeta d_1$ accounts for the corresponding cost or deviation. A collision or unsuccessful action ($\dot{o}_i^t=C$) results in a negative normalized reward, whereas no reward is assigned when $\dot{o}_i^t\in\{B,I\}$. Accordingly, the reward function is defined as:}

\begin{equation}\label{user_reward}
\small
{\rho}_{i}^{t} = 
\left\{\begin{array}{ll}
    r_{i}^{t} / R_{max} \ - \alpha \max_{j}{\bar{v}_{j}^{t}} & \mbox{if} \; \makecell[l]{\dot{o}_{i}^{t} = S \\ \land \ (\mathcal{U}^{t} = \mathcal{F}_{\alpha}^{t})} \\
    r_{i}^{t} / R_{max} \ - \zeta d_{1} & \mbox{if} \; \makecell[l]{\dot{o}_{i}^{t} = S \ \land \\ (\mathcal{U}^{t} = \mathcal{E}_{\zeta, d_{1}, d_{2}}^{t})} \\
    - r_{i}^{t} / R_{max} & \mbox{if} \; \dot{o}_{i}^{t} = C\\
    0 & \mbox{if} \; \dot{o}_{i}^{t} \in \{ {{B, I}} \}
\end{array}\right. \notag
\end{equation}

\begin{equation}\label{reward}
\boldsymbol{\rho}^{t} = \frac{1}{\mathcal{N}} \ \sum_{i \in \mathbb{N}} {{\rho}_{i}^{t}}
\end{equation}

\textcolor{black}{It should be emphasized that a collided macro-action is re-decided rather than retransmitted. Following the collision penalty $-r_{i}^{t}/R_{max}$ defined above, the next decision is a fresh selection over the entire action space in \eqref{action_space}, so the UE may re-attempt immediately, sense first, switch channel, or shorten its packet, and the timing, channel, and length of a retry are learned from \eqref{state_space} instead of being fixed by an ARQ timer and a backoff rule. Complementarily, since a segment is delivered whenever any associated UE succeeds, per \eqref{segment_indicator}, spatial semantic redundancy acts as an implicit repetition code across UEs rather than across successive attempts of the same UE, which is precisely what the assisted throughput $\ddot{x}_{i}^{t}$ in \eqref{ue_throughput_decomposed} and the assisted transmission $\dddot{o}_{i}^{t}$ quantify. Explicit retransmission with soft combining would add a third, temporal form of redundancy and is deferred to future work.}

\subsubsection{Training Process}

Our proposed methodology lies between the Value Decomposition Network (VDN) framework \cite{sunehag2017value} and the QPLEX algorithm \cite{wang2020qplex}, thus adhering to the Centralized Training and Decentralized Execution (CTDE) paradigm. This strategy operates on the premise that during the training phase, full access to the global system state is available, whereas user decisions are made based on their local information. Similarly, the training process of our proposed method is centralized within the SBS, where UE policies are being trained. Given that the SBS possesses knowledge of UE experiences, including their actions and observations, there is no need for agents to transmit their experiences to the SBS. However, it is assumed that the SBS periodically updates and disseminates policies to the UEs via high-bandwidth and low-latency dedicated communication channels. UEs autonomously select their actions based on their respective policies and individual observations without relying on the actions or states of other UEs.

To implement the VDN, we employ D3QL, as detailed in \cite{wcnc_2024}. In this approach, the expected value of state-action pairs for each UE is represented as the combination of state and advantage values, specifically  $Q_{i} = V_{i} + {adv}_{i}$, where  $Q_i$ is the Q-value,  $V_i$  is the state value, and  ${adv}_{i}$ is the advantage value. This methodology enhances the stability and performance of the learning process by clearly separating the state value from the advantage of specific actions, allowing for more accurate value estimation and decision-making. \textcolor{black}{The overall action-value function is obtained by summing the individual UE action-value functions, enabling the SBS to train all agents using the common system reward defined in \eqref{reward}. This is expressed as:}

{\color{black}
\begin{align}\label{training}
Q_{tot}(\boldsymbol{S}^{t}, \boldsymbol{\mathcal{A}}^{t})
&= \sum_{i \in \mathbb{N}^{t}_{+}} Q_i(s_i^{t},a_i^{t})
\end{align}

Algorithm~\ref{PRISM} summarizes the interaction between decentralized execution and centralized training. In Lines 4--16, each active UE applies an $\epsilon$-greedy policy to either select the locally optimal action or explore a random sensing or transmission action. After executing the selected action, the UE receives the required feedback from the SBS and constructs its next local state. In Lines 18--23, the SBS computes the common system reward, stores the joint experience in the replay memory, samples a training batch, and updates the UE networks using the centralized value ($Q_{\mathrm{tot}}$). The updated network parameters are then disseminated to the UEs for subsequent decentralized decision-making. The exploration probability is initialized as $\epsilon=1$ and is updated according to ($\epsilon \gets \epsilon \epsilon'$), where $\epsilon'$ is the multiplicative decay factor. This decay continues until $\epsilon$ reaches the minimum exploration probability ($\widetilde{\epsilon}$).
}

\begin{algorithm}[t!]
\small
\caption{PRISM}\label{PRISM}
\KwIn{$\mathcal{T}$, $\epsilon'$, and $\widetilde{\epsilon}$}
$\boldsymbol{\mathcal{W}} \leftarrow \mathbf{0}$, $\boldsymbol{\mathcal{W}}^{-} \leftarrow \mathbf{0}$, $\epsilon \gets 1$, $memory \gets \{\} $\\
\ForEach{$t$ in $\{0, \ldots, \mathcal{T} \}$}
{
    \textcolor{gray}{$\star$ Decentralized Execution (4-16)}  \\
    \ForEach{$i$ in $\mathbb{N}^{t}_{+}$}
    {
        $\iota \gets$ generate a random number from $[0:1]$ \\
        \If{$\iota > \epsilon$}
        {
            $(r_{i}, c_{i}) \gets$ argmax$_{\rho \in \boldsymbol{a}_{i}} Q({s}_{i}^{t},  \rho, {\mathcal{W}}_{i})$ \\
        }
        \Else
        {
            select a random $(r_{i}, c_{i})$ from $\boldsymbol{a}_{i}$
        }
        \If{$r_{i} > 0$}
        {
            start transmitting a packet with length $r_{i}$ \\
            on channel $c_i$
        }
        \Else
        {
            sense channel $c_i$
        }
        \textcolor{black}{obtain $\dot{o}_{i}^{t}\!\in\!\{B,I\}$ locally after sensing; receive $\bar{v}_{i}^{t}$, $\ddot{o}_{i}^{t}$, $\dddot{o}_{i}^{t}$ and, after transmission, $\dot{o}_{i}^{t}\!\in\!\{S,C\}$ from the SBS} \\
        construct ${s}_{i}^{t+1}$ \\
        
    }
    \textcolor{gray}{$\star$ Centralized Training (18-23)} \\
    calculate $\boldsymbol{\rho}^{t}$ according to \eqref{reward} \\
    $memory \gets \{ \boldsymbol{\rho}^{t} \} \cup \{({s}_{i}^{t}, (r_{i}, c_{i}), {s}_{i}^{t+1}) \big| \ i \in \mathbb{N}^{t}_{+} \}$ \\
    choose a batch of samples from $memory$\\
    train the agent according to \eqref{training} \\
    \If{$\epsilon > \widetilde{\epsilon}$}
    {
        $\epsilon \gets \epsilon \epsilon'$
    }
}
\end{algorithm}

%% file: sections/5_evaluation.tex
\section{Evaluation}\label{S_EVA}

Two main scenarios were carried out to evaluate the performance of the PRISM scheme, one dedicated to $\alpha$-fairness maximization (problem $P_{1}$), and another to energy efficiency maximization (problem $P_{2}$). \textcolor{black}{The experiments are designed to answer three specific questions: (i) what is gained by adding semantic awareness to a modern MADRL-based MAC policy, (ii) how far a decentralized policy is from a centralized oracle under the same system model, and (iii) how the gain changes with the amount of inter-user semantic overlap. The association matrices used for the evaluations are synthetic stress-test instances for these questions; they are not presented as a community benchmark or as a substitute for application-level datasets.}

{
\color{black}
Our principal learning baseline is PRISM-lite, a literature-aligned semantic-oblivious MADRL policy. Contemporary learning-based MAC schemes commonly combine decentralized decisions from local histories with cooperative multi-agent learning and collision feedback \cite{naparstek2018deep, guo2022multi, miuccio2024learning}. PRISM-lite follows these general design principles and uses the same CTDE procedure, value-decomposition and double-dueling Q-learning backbone, history representation, channel--packet-length action space, reward, model capacity, and training budget as PRISM. Its only substantive difference is that the self- and assisted-semantic-transmission observations are removed. Thus, PRISM-lite is not intended to imitate a particular published protocol. It is a strong matched baseline that instantiates the established semantic-oblivious MADRL approach in the proposed asynchronous, variable-packet-length environment. Holding the learning and MAC components fixed avoids attributing differences caused by another algorithm, action space, or signaling model to semantic awareness.

\textcolor{black}{CB denotes the model-specific collision-free centralized benchmark used in the evaluation}, RND is a sanity check, and ORTH is an idealized semantic-orthogonality capability reference motivated by MDMA \cite{zhang2023model}, O-MDMA \cite{liang2024orthogonal}, and ToDMA \cite{qiao2025token}. These published schemes show that signals carrying separable semantic content can, under method-specific encoders and decoders, share transmission resources. ORTH abstracts that common capability by allowing collision-free concurrent reception when the transmitted segments do not overlap. This deliberately favorable rule isolates the potential benefit of semantic orthogonality without claiming to reproduce a particular encoder, decoder, modality, or reconstruction metric. Accordingly, ORTH is not a measured implementation or formal upper bound of those methods; it is a best-case, model-level reference against which the distinct benefit of exploiting semantic similarity can be assessed. Optimal throughputs are derived from a centralized scheduler that allocates time slots and channels based on optimal solutions computed \textcolor{black}{by the Gurobi optimizer.}
}

In both scenarios, we set $\mathcal{N} = 8$, $\mathcal{C} = 4$, ${R}_{max} = 5$, and $\mathcal{T} = 30000$. \textcolor{black}{Notably, across all experiments, a $10\%$ probability of channel failure per time frame is incorporated to preserve the realism of wireless network assumptions. However, the dominant performance limitation in this study is due to collisions among users.} Furthermore, similar to investigating optimal solutions, to demonstrate the impact of semantic sharing among UEs, we define and utilize a set of UE-segment association square matrices $A_{\mathcal{N}, q}$ in both scenarios. In this section, $A_{\mathcal{N}, q}$ translates to a non-diagonal relation that occurs with probability $q$, while a diagonal relation occurs with probability $1$. Table \ref{tab_sim_par} contains hyperparameters and configurations of simulations. Notably, our three-layer fully-connected neural network has a Multiply-ACCumulate (MAC) value of $\approx 45$k, indicating its status as a comparatively compact model. \textcolor{black}{On our simulation platform\footnote{\color{black} An NVIDIA RTX 4000 Ada Generation (20 GB) and an Intel(R) Core(TM) Ultra 9 285K (3.70 GHz) CPU}, each training run consisting of 30,000 time steps requires approximately 70–90 minutes to complete. This duration includes the forward pass, backward pass, and memory replay.} \textcolor{black}{Finally, each association matrix is fixed within a run and varied across runs; hence, the evaluation assumes stationary association statistics during learning. Instantaneous semantic contributions still vary with $\delta_k^t$ and access outcomes. Adaptation to within-run drift is left for future work.}

\begin{table}[t!]
\caption{Training Configuration.}
\begin{center}
\small
\begin{tabular}{|l|l|}
\hline
\multicolumn{1}{|c|}{\textbf{Parameter}} &
  \multicolumn{1}{c|}{\textbf{Value}} \\
\hline
Short-term history size ($\mathcal{H}_{S}$) & $5$ experiences \\ \hline
Long-term history size ($\mathcal{H}_{L}$) & $20$ experiences \\ \hline
Capacity of experience memory & $1000$ experiences \\ \hline
Batch size & $64$ \\ \hline
Discount factor ($\gamma$) & 0.9 \\ \hline
Learning rate & $0.001$ \\ \hline
Exploration parameters $\widetilde{\epsilon}$, $\epsilon'$ & 0.005, 0.999 \\ \hline
Approximator model & \begin{tabular}{@{}c@{}}A fully-connected network \\ with $256$, $128$ and $64$ units  \end{tabular}  \\ \hline
Training frequency & \begin{tabular}{@{}c@{}} Every $1$ step \end{tabular}  \\ \hline
\begin{tabular}{@{}c@{}} Target network \\ update frequency \\ \end{tabular} & Every $50$ steps \\
\hline
\end{tabular}
\label{tab_sim_par}
\end{center}
\end{table}

\subsection{$\alpha$-fairness Maximization}
\begin{figure}[t!]\centering
\centerline{\includegraphics[width=3.6in]{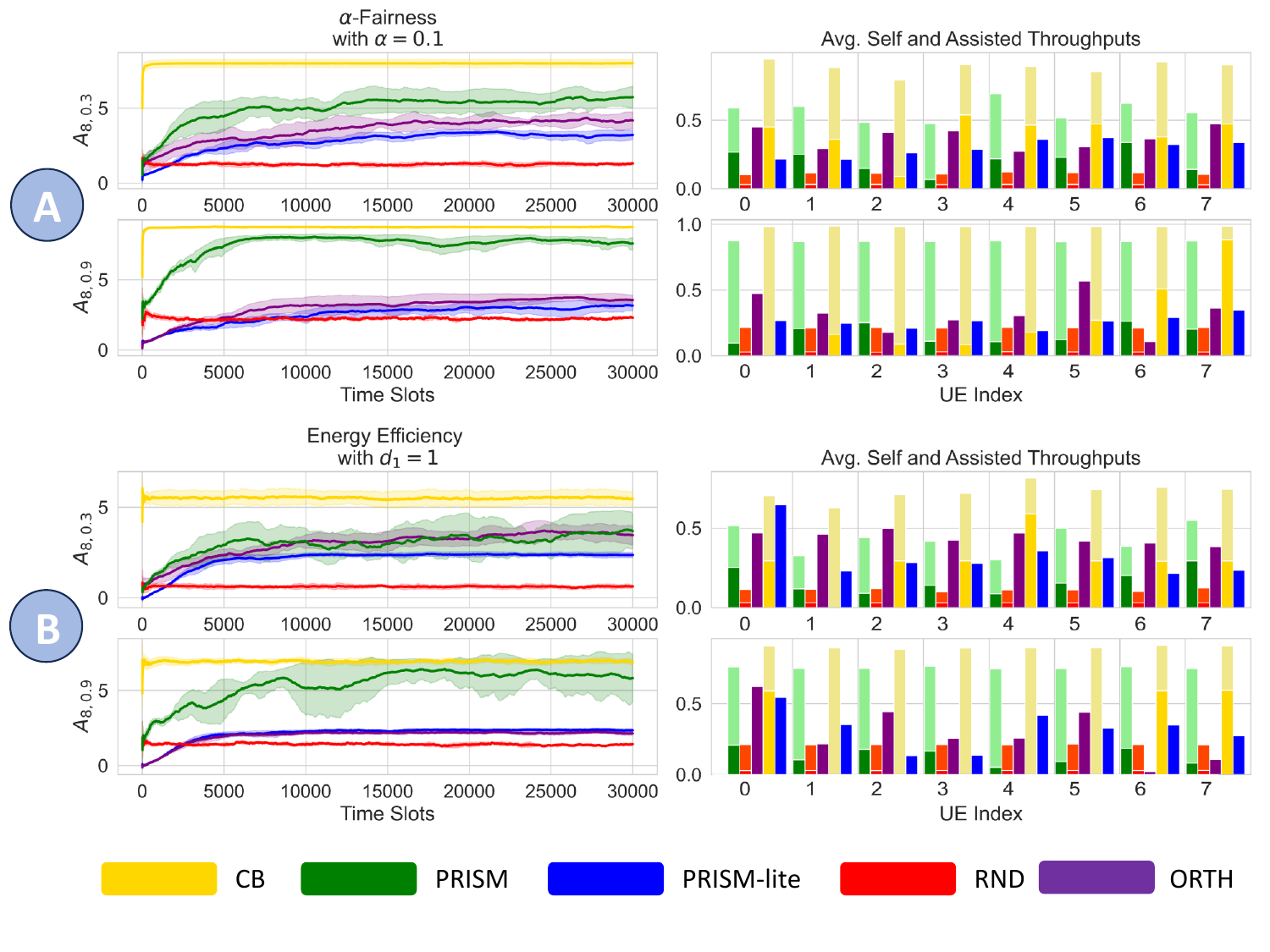}}
\vspace{-10pt}
\caption{Objective functions over time and all-time average UE throughputs, encompassing selected configurations of experiments A and B, comparing \textcolor{black}{CB}, PRISM, PRISM-lite, RND, and ORTH. Each line represents the average of five executions with random channel impairments, with shaded areas indicating the standard deviation.
It is noteworthy that, when present, the lighter segment of each UE throughput signifies the assisted throughput for that particular UE. }
\label{Time_based}
\end{figure}

In this scenario, we compare algorithms based on their ability to maximize the fairness-utilization trade-off, quantified using the \(\alpha\)-fairness objective function defined in \eqref{alpha_fairness}. Experiments are conducted for \(\alpha\) values of 0 (i.e., sum-rate maximization), 0.1, 0.5, {\color{black} 1 (Proportional Fairness), 2, and $\alpha \to \infty$ (max-min fairness)}. In Fig. \ref{Time_based}-(a), the evolution of \(\alpha\)-fairness during the training procedure of 8 UEs over time is depicted for \(\alpha = 0.1\) on the left. On the right, the subfigure shows the all-time average self and assisted throughputs for each UE. The results compare \(A_{8, 0.3}\) (little shared semantics among users) and \(A_{8, 0.9}\) (high rate of semantic sharing among users). When there is little shared semantic information to exploit (e.g., users’ non-overlapping observations), there is less benefit in incurring the additional complexity of semantic orchestration. On the bright side, when users’ streams are highly interdependent, semantic-aware multiple access can approach optimal performance in both scenarios.

\begin{figure}[t!]\centering
\centerline{\includegraphics[width=3.5in]{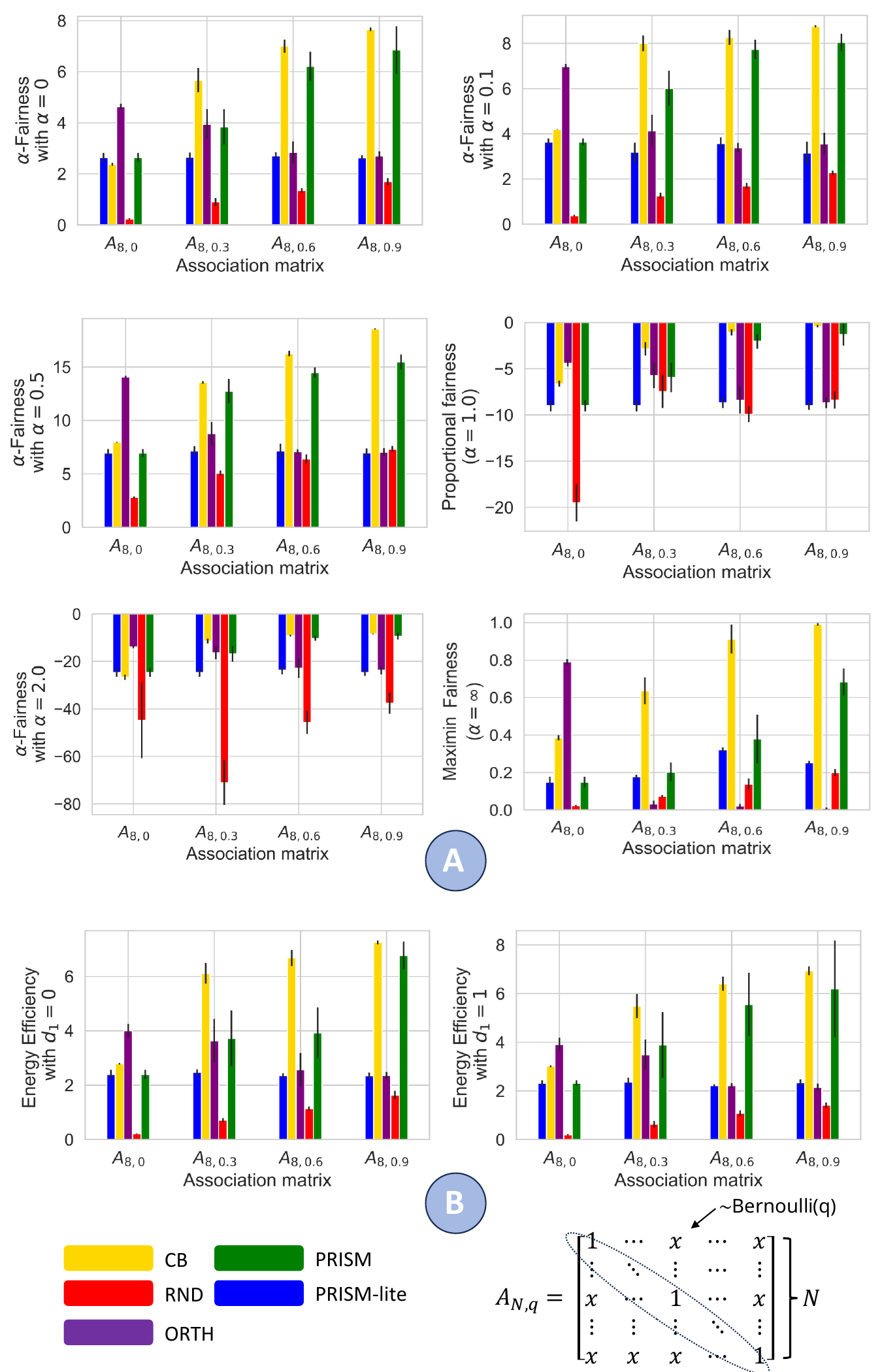}}
\caption{\textcolor{black}{Average of the last 1000 time slots of objective functions, encompassing experiments A and B, comparing \textcolor{black}{CB}, PRISM, PRISM-lite, RND, and ORTH. The reported results are based on averaging the outcomes of five rounds of simulations with random channel impairments, with error bars indicating the standard deviation.}}
\label{aggregated}
\end{figure}

Aggregated results for all values of \(\alpha\) and association matrices are presented in Fig. \ref{aggregated}-(a), illustrating that PRISM outperforms both random and PRISM-lite approaches, approaching optimal performance in many scenarios. It also outperforms the upper bound of orthogonal approaches in environments with a high shared semantics rate. For \(\alpha = 0\) (sum-rate maximization), the focus is on maximizing total throughput without considering fairness, potentially allowing some UEs to monopolize the channel. As \(\alpha\) increases to 0.1 \textcolor{black}{and beyond}, the trade-off shifts towards more equitable throughput distribution among UEs. The comparison between \(A_{8, 0.1}\) and \(A_{8, 0.9}\) demonstrates that higher rates of semantic sharing lead to better performance in terms of both self and assisted throughputs.

The results suggest that PRISM successfully balances fairness and utilization, with greater capability observed for lower \(\alpha\) values. This is likely due to the inherently cooperative nature of fairness, making it challenging to achieve in a distributed multiple-access setup. The incorporation of semantic sharing among UEs significantly enhances the system's performance, especially in terms of fairness and utilization trade-offs. This highlights PRISM's capability to manage resources efficiently in a distributed multiple-access environment, demonstrating superior performance and approaching optimal outcomes.

{\color{black} The comparison with ORTH further illustrates the difference between semantic orthogonality and semantic similarity. ORTH can be competitive when \(q\) is small because non-overlapping segments are common, making its idealized collision-free concurrent reception assumption favorable. As \(q\) grows, fewer transmissions satisfy the non-overlap condition, while PRISM benefits directly from the increased semantic commonality.}

\subsection{Energy Efficiency Maximization}
In this scenario, we evaluate the algorithms based on their ability to maximize overall energy efficiency, as defined in \eqref{energy_utility_normalized}. The parameters \(\zeta\) and \(d_{2}\) are set to 0.1 and 0.01, respectively. However, \(d_{1}\) is assigned two values: 0 to exclude semantic transcoding costs, and 1 to include them. More detailed results are illustrated in Fig. \ref{Time_based}-(b) for \(d_{1} = 1\). The subfigure on the left shows the evolution of energy efficiency during the training procedure of 8 UEs over time, while the right side displays the all-time average self and assisted throughputs for each UE. The results compare \(A_{8, 0.3}\) and \(A_{8, 0.9}\), similar to the previous scenario. Aggregated results for all values of \(d_{1}\) and association matrices are depicted in Fig. \ref{aggregated}-(b).

The figures illustrate that PRISM outperforms both random and PRISM-lite approaches, approaching optimal performance in terms of energy efficiency. For both \(d_{1} = 0\) and \(d_{1} = 1\), the energy efficiency shows significant improvements, indicating the effectiveness of PRISM in managing energy consumption, even when semantic transcoding costs are included. The comparison between different association matrices demonstrates that higher semantic sharing rates lead to improved performance. These results highlight PRISM's capability to manage resources efficiently in a distributed multiple-access environment, particularly in terms of energy efficiency. Moreover, compared to orthogonal approaches, PRISM performs better in scenarios with higher semantic sharing rates, promising that these approaches can be used complementarily. 

\textcolor{black}{It is worth noting that the two scenarios optimize and report $\mathcal{F}_{\alpha}^{t}$ and $\mathcal{E}_{\zeta,d_{1},d_{2}}^{t}$ separately; they do not cross-evaluate each trained policy. Optimizing one objective therefore does not guarantee the other. Nevertheless, the two objectives are aligned in the direction that matters here: both reward the same utilization term and differ only in the penalized quantity, i.e., throughput dispersion under $\mathcal{F}_{\alpha}^{t}$ versus transmission and inference energy under $\mathcal{E}_{\zeta,d_{1},d_{2}}^{t}$, while assisted throughput relieves both penalties at once, as it raises a UE's throughput with no additional energy cost and lifts precisely the UEs that did not transmit successfully. Consistently, the ordering of the schemes and the growth of the semantic gain with $q$ reported above were observed under both objectives. Cross-evaluation and a scalarized fairness--energy Pareto analysis are left for future work.}

\color{black}
\subsection{Sensitivity to the Number of UEs and Channels}
\begin{figure}[t!]\centering
\centerline{\includegraphics[width=3.5in]{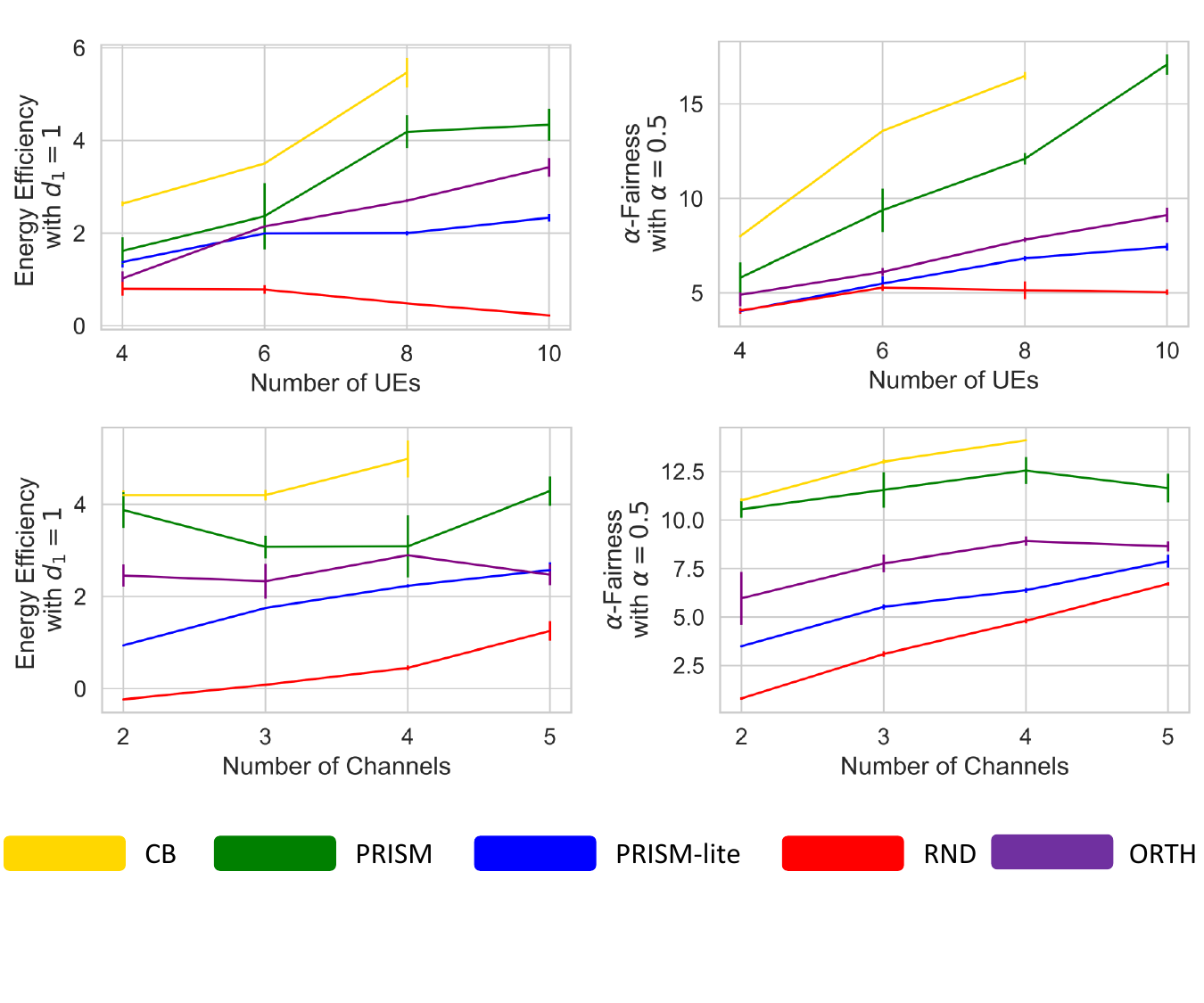}}
\caption{\textcolor{black}{Sensitivity of energy efficiency and \(\alpha\)-fairness to the number of UEs and channels, comparing \textcolor{black}{CB}, PRISM, PRISM-lite, RND, and ORTH. The top row varies the number of UEs, and the bottom row varies the number of channels. The reported results are averaged over five executions with random channel impairments, with error bars indicating the standard deviation. For some cases, the optimal values could not be obtained within a reasonable computational time.}}
\label{num_user_num_channel}
\end{figure}

Fig. \ref{num_user_num_channel} evaluates whether the observed behavior persists when the network size changes. The top row varies the number of UEs, and the bottom row varies the number of channels. The left column reports energy efficiency with \(d_{1}=1\), while the right column reports \(\alpha\)-fairness with \(\alpha=0.5\).

As the number of UEs increases, PRISM benefits from the larger pool of potentially related semantic segments. The widening performance gap between PRISM and PRISM-lite across both objectives demonstrates that semantic awareness becomes increasingly valuable under heavier contention, rather than constituting an advantage limited to small-scale networks. Moreover, the evaluation across varying numbers of users reveals that ORTH improves at a slower rate than PRISM, since its gains depend on semantic non-overlap, whereas PRISM can exploit the increasing number of overlapping segments. Increasing the number of channels reduces the occurrence of collisions, thereby improving the performance of PRISM-lite and RND, particularly in terms of \(\alpha\)-fairness. Nevertheless, PRISM remains the strongest decentralized method, as it jointly benefits from the collision reduction enabled by additional channels and the assistance provided by semantic information. Regarding energy efficiency, the evaluation across varying numbers of channels does not exhibit strictly monotonic behavior at every point. In particular, when \(d_{1}=1\), additional transmission opportunities may also incur further transcoding costs. This observation highlights that PRISM addresses a joint communication-computation trade-off, rather than merely maximizing the number of successful transmissions.

\color{black}

%% file: sections/6_conclusion.tex
\section{Conclusion}\label{S_CON}

In this paper, we presented a novel framework for distributed multiple access that capitalizes on the revolutionary potential of semantic communication. By explicitly modeling the correlation among user source data—a critical feature of emerging 6G applications like the Metaverse and holographic telepresence—we established a foundation for optimizing spectral usage. We formulated two distinct optimization problems targeting the fairness-utilization trade-off and energy efficiency, deriving the PRISM (Protocol for Redundancy Identification in Semantic Multiple-access) scheme to solve them. Our evaluation demonstrates that integrating semantic intelligence into distributed MAC decisions significantly enhances network sustainability compared to semantic-oblivious baselines.

Looking ahead, we aim to extend this framework to scenarios involving dynamic or imperfectly characterized association matrices, reflecting the fluid nature of real-time user data. \textcolor{black}{This includes UEs joining or leaving the deployment, or changing their transmission and semantic-generation patterns during operation, which breaks the stationarity assumed in the Evaluation section and calls for online adaptation of the learned policies.} We propose leveraging Continual Learning (CL) algorithms to handle these transitions, ensuring forward and backward transfer as user-segment associations evolve. Furthermore, we intend to investigate the granularity trade-off inherent in semantic segmentation: fine-grained segmentation offers precision at the cost of complexity, while coarse granulation risks association errors. \textcolor{black}{We also intend to incorporate more realistic wireless channel models, including time-correlated fading and dynamic channel quality estimation, to better capture practical communication environments.} \textcolor{black}{In the same direction, we plan to relax the error-free control-feedback assumption of Section~\ref{S_SES} and to incorporate explicit retransmission mechanisms, such as HARQ with soft combining, into the macro-action structure via a redundancy-version selector and an SINR-based capture model, so that a partially decodable transmission is no longer treated as a total loss.} Finally, we plan to validate the PRISM framework by deploying it within a real-world testbed for one of the identified 6G use cases.